\documentclass[article]{aastex631}
\usepackage{graphicx}
\usepackage{amsmath}
\usepackage{amssymb}
\usepackage{natbib}

\def\apjl{Astrophysical Journal, Letters}
\def\apjs{Astrophysical Journal, Supplement}
\def\aap{Astronomy and Astrophysics}

\shorttitle{Understanding  periodicities}
\shortauthors{Zharkova\& et al}

\begin{document}
 
\title{Periodicities of solar activity and solar radiation derived from observations and their links with the terrestrial environment}

\author{Valentina V. Zharkova}
\affil{Department of MPEE, University of Northumbria, Newcastle upon Tyne,  UK}
\affil{ZVS Research Enterprise Ltd., London, UK} 
\email{valentina.zharkova@northumbria.ac.uk}
\author{Irina Vasilieva}
\affil{Department of Solar Physics, Main Astronomical Observatory, Kyiv,  Ukraine}
\affil{ZVS Research Enterprise Ltd., London,  UK}
 \author{Simon J. Shepherd}
\affil{PRIMAL Research Group, Sorbonne Universite, Paris,  France}
\affil{ZVS Research Enterprise Ltd., London, UK} 
\author{Elena Popova}
\affil{Centro de Investigación en Astronomía, Universidad Bernardo O’Higgins, Santiago, Chile}
 \begin{abstract}
Solar magnetic activity is expressed via variations of sunspots and active regions varying on different timescales.  The most accepted is an 11-year period supposedly induced by the electromagnetic solar dynamo mechanism. There are also some shorter or longer timescales detected: the biennial cycle (2-2.7 years), Gleisberg cycle (80-100 years), and Hallstatt's cycle (2100-2300 years).  Recently, using Principal Component Analysis (PCA) of the observed solar background magnetic field (SBMF), another period of 330-380 years, or Grand Solar Cycle (GSC), was derived from the summary curve of two eigenvectors of SBMF.
In this paper, a spectral analysis of the averaged sunspot numbers, solar irradiance, and the summary curve of eigenvectors of SBMF was carried out using Morlet wavelet and Fourier transforms. We detect a  10.7-year cycle from the sunspots and modulus summary curve of eigenvectors as well a 22 years cycle and the grand solar cycle of 342-350-years from the summary curve of eigenvectors. The Gleissberg centennial cycle is only detected on the full set of averaged sunspot numbers for 400 years or by adding a quadruple component to the summary curve of eigenvectors. Another period of 2200-2300 years is detected in the Holocene data of solar irradiance measured from the abundance of $^{14}$C isotope. This period was also confirmed with the period of 2100 years derived from a baseline of the summary curve, supposedly, caused by the solar inertial motion (SIM) induced by the gravitation of large planets. The implication of these findings for different deposition of solar radiation into the northern and southern hemispheres of the Earth caused by the combined effects of the solar activity and solar inertial motion on the terrestrial atmosphere are also discussed
  \end{abstract}

\keywords{Sun: magnetic field -- Sun: solar activity -- Sun: sunspots  -- Sun: solar dynamo}

\section{Introduction} \label{intro}
Currently, solar magnetic activity is expressed by the number of sunspots on the solar surface, by energetic events in the solar atmosphere, or by variations of various radioisotopes in the terrestrial biomass, which vary on a wide range of timescales. Most prominent is the 11-year solar cycle, which is significantly modulated by some longer timescales of a few hundred or a few thousand years. Solar activity is debated for decades to be linked to the variations of terrestrial temperature \citep{Eddy1976, Connolly2021} recently objected by IPCC. Understanding the nature of solar variability and the effects of the Sun on the terrestrial environment is the essential precondition for an attempt to predict future solar activity and terrestrial variation levels not only on 11 years but also on a longer timescale.

Furthermore, the frequencies derived from averaged sunspot numbers are shown by \citet{Cameron2017} to depend on the data they used: 1) sunspot data for the past 370 years and 2) the  carbon isotope $^{14}$C data 2 produced by a number of authors \citep[see, for example,][and references therein]{usoskin2016, usoskin2021}. Data 1 does not contain any other data than those varying with 11-years period, as its size  the full data 1 duration is within the  size of a single grand solar cycle. The other frequencies of 93 and 200 years derived from sunspot index \citep{Cameron2017, Cameron2019} are assigned to the normal noise, although noting that these frequencies might be real.  The carbon data 2 is strongly polluted by various other processes, e.g. a change of the local background intensity, natural conditions destroying the biomass, etc. that their conclusions should be taken with care \citep{Brehm2021}.

Systematic properties of sunspot groups \citep{hale1919}) indicate that they originate in the solar convection zone from a reservoir of the  East-West orientated ( toroidal) magnetic field. This magnetic field is generated by winding up a poloidal magnetic field (such as a dipole magnetic field aligned with the sun's rotation axis) by the differential rotation of the Sun so that its axisymmetric component dominates. The poloidal field is (re) generated against the effect of Ohmic decay by the collective effect of loops formed from the toroidal field by convective flows and/or magnetic buoyancy. The loops become twisted owing to the Coriolis force and thus acquire a systematic meridional component \citep{parker55, babcock61}.  This interaction of the toroidal and poloidal magnetic fields leads to a 22-year magnetic cycle and an 11-year cycle of sunspot activity \citep{parker55, babcock61}.

\citet{jones10} presented a detailed description of the four main models of the mean-field dynamo starting from the first one, Interface Dynamo,  suggested by \citet{parker55}, which was later modified by \citet{parker93} suggesting that the shear occurs in the tachocline and the $\alpha$-effect in the convection zone. This separation is desirable because then the strong fields in the tachocline might not suppress the $\alpha$--effect in the convection zone so much, and the tachocline is known to be the region of strong shear. 

Second model, Flux Transport Dynamos, which goes back to the ideas of \citet{babcock61} and \citet{leighton69}, and are similar to the interface dynamo concept, in that they are $\alpha$- dynamos, in which the shear is in the tachocline while the $\alpha$--the effect is in the convection zone that can explain the observed twisting of a field in line with Joy’s law. So far, this is quite similar to the interface dynamo, but the new feature of flux transport dynamos is that the poloidal flux is transported back to the tachocline by a meridional circulation rather than by diffusion alone \citep{Choudhuri1995, Dikpati1995}. 

The third model,  Tachocline Based Dynamos, which has proposed \citep[e.g.][]{Spiegel1980} that the solar dynamo generation occurs in the tachocline and lower convection zone only. On this view, what one sees on the surface is merely a rather scrambled view of this process as the generated field rises through the convection zone to the surface. Low-order models demonstrate how nonlinear interactions can lead to the modulation of cyclic activity and solar luminosity. 

These models may be constructed in two ways. The first involves the truncation of the relevant partial differential equations (in this case either the mean field equations or the full MHD equations). For nonlinear interactions of the magnetic field with the differential rotation, the dynamics reduces to a complex generalisation of the Lorenz equations \citep{Jones1985}. If the magnetic field acts back on a dynamic alpha-effect then the structure of the low-order model is different \citep{Covas2001}. 

An alternative way is to derive low-order models using normal-form theory and the underlying symmetries of the equations. This has the advantage that the dynamics found is structurally stable and robust (at least close to the initial bifurcation). Global Computational Models \citep[see, for example,][]{Miesch2009, Brun2009} are computational models, which are constructed to examine the generation of large-scale modes in a solar context. Computational models, thoough, have significant difficulty in reaching the correct parameter regime not only in the momentum equation  but also for the induction equation itself. 

There are also a number of papers describing the variability of solar activity with electromagnetic dynamo theory, which can be roughly divided into two classes: (1) nonlinear dynamics and deterministic chaos, and (2) random fluctuations of dynamo excitation. The nonlinear dynamics approach typically considers the bifurcation structure of low-order dynamical systems (see reviews by \citet{Weiss1990, Tobias2002} but models based on nonlinear turbulent PDEs have also been frequently investigated \citep[e.g.][]{Schmitt1989, Charbonneau2010, Tobias11, Charbonneau2014}.

Hence, in general, dynamo theory produced by a dipole magnetic field is divided into two classes: large-scale and small-scale dynamo models \citep{jones10, Nigro2017}.  In \citet{Nigro2017} the authors consider kinematic dynamo action in a sheared helical flow at moderate to high values of the magnetic Reynolds number (Rm). They find exponentially growing solutions that, for large enough shear, take the form of a coherent part embedded in incoherent fluctuations. They further argue that although the growth rate is determined by small-scale processes, the period of the coherent structures is set by mean-field considerations.
It is obvious that some stochastic component in solar activity may possibly be present in the solar dynamo, as it is natural for large complex dynamical systems while a ratio between non-stochastic and stochastic components of solar activity are still debated depending on the model approaches. 

Recently,  \citet{zharkova12, zhar15} suggested using an additional proxy of solar activity - the eigenvectors of the solar background (poloidal) magnetic field  (SBMF), obtained from the  low resolution synoptic magnetic maps of Wilcox Solar Observatory. These eigen vectors  consider the low-magnitude solar background magnetic field and are not contaminated by strong magnetic fields of magnetic loops and sunspots \citep{Zharkova2022}, which are present in the high-resolution synoptic magnetic maps of Kitt Peak Observatory \citep{Lawrence04, Cadavid2005}.  By applying the principal component analysis (PCA) to the synoptic magnetic maps for cycles 21-23  \citep{zharkova12} and, recently, for 21-24 \citep{Zharkova2022} the authors identified and confirmed a number of eigenvalues and eigenvectors from the SBMF representing magnetic waves of the solar surface. The eigenvectors, or magnetic waves, are found to appear in pairs assumed to be generated by different magnetic sources in the solar interior, e.g. by a dipole, quadruple, sextuple sources, etc. 

The first pair of eigenvectors covered by the largest amount of the magnetic data by variance, or principal components (PCs), reflects the primary waves generated by the solar dynamo produced by the dipole magnetic sources \citep{zhar15}. The temporal features of the summary curve of these two PCs show a remarkable resemblance to the sunspot index of solar activity (representing a toroidal magnetic field of active regions) for cycles 21-23 \citep{zhar15} and cycles 21-24 \citep{Zharkova2022}. This correspondence occurs despite the PCs and sunspot indices representing very different entities of solar activity: poloidal magnetic field for PCs and toroidal magnetic field for sunspot numbers.  However, their similarity and periodicity allowed  \citet{zhar15} to suggest using this summary curve of the two PCs of SBMF as a new, or additional, solar activity proxy. 

The advantage of using the summary curve as a solar activity proxy instead of the averaged sunspot numbers is the use of the eigenvectors of the solar (poloidal) magnetic field oscillations,  derived from the surface magnetic field measurements on the whole solar disk which are expressed then by mathematical formulae and, thus, can be extrapolated for any time. The ample magnetic full disk data significantly reduces the errors in the magnetic wave definition and introduces an extra-parameter, a leading polarity of SBMF \citep{zhar15},  which is shown to be in anti-phase with the magnetic polarity of leading sunspots \citep{zharkov08}.   

The two largest eigenvectors,  or  PCs, are the two different magnetic waves, which have amplitudes and phase shifts varying with time and latitudes. As a result, the summary curve of the resulting magnetic wave has some elements of stochasticity induced by the wave interference, caused by amplitude variations and the phase difference producing the envelope period of amplitude oscillation about 330-400 years called a grand solar cycle separated by grand solar minima \citep{zhar15, zhar2020}. The phase shift between these two dipole waves changes the magnetic field shapes within 22-year cycles produced during grand solar cycles (GSC) \citep{Zharkova2022b}. The addition of the quadruple magnetic field can contribute to the formation of the centennial (Gleisberg) solar activity cycle \citep{Popova18} and can be directly linked to a flaring activity \citep{Zharkova2022}.

This new proxy of solar activity is essential progress in understanding the nature of solar activity because: 1) it considers directly a poloidal magnetic field used in dynamo models; 2) it derives that magnetic waves visible on the solar surface are the superposition of the waves created by dipole, quadruple, sextuple and octuple magnetic sources; 3) it detected that the dynamo waves are formed in pairs, or in two different layers, with the first pair data (67\% by STD) linked to dipole magnetic sources; 4) superposition of these waves defines the observed magnetic waves on the solar surface.

Knowing this specific effect of the phase shifts, it makes much easier to model dynamo waves with similar phase shifts in each layer, whose superposition displays visible stochasticity without any heavy assumptions on other conditions in the solar interior as was required in the paper by \citet{Cameron2017, Cameron2019} where stochastic dynamo equations were applied to two layers in the solar interior, instead of the normal mean dynamo models with meridional circulation. 
 
Therefore, we need to clarify to what extent randomness, intrinsic periodicities apart from the 11-year cycle, and nonlinearities of the underlying dynamo process generating the solar magnetic field contribute to the observed long-term variability of solar activity. In fact, in the current paper, we will investigate the frequencies of solar activity derived from different datasets currently available. This analysis definitely indicates that the sunspot data has not only 11 but 22-year cycles. The latter makes much more sense, given the fact that the leading polarity change in sunspots is 22 years and not 11.

In addition,  the analysis of variations of carbon $^{14}C$ isotope abundance  \citep{Reimer09} and irradiance oscillations over the past 12 000 years \citep{vieira2011, vieira2011Cat}  clearly demonstrate the well-defined millennial oscillations with a period of 2200 years \citep{Steinhilber12, vieira2011}, 2300 years \citep{Scafetta2016} or 2400 years \citep{usoskin2016}. These variations constitute the two-millennial Hallstatt's cycle of solar irradiance variations \citep{Scafetta2016}.

This Hallstatt cycle is also detected in the magnetic field baseline oscillations with the current  Hallstatt's cycle to be started at the Maunder minimum and progress in ascending phase until 2500-2600 meaning the increase of solar irradiance during this period. Hence, this two-millennial period of solar irradiance oscillation combined with the grand solar cycle of 330-400 years, Gleisberg cycles of 80-100 years, and the basic cycles of solar activity of 11/22 years are the main ingredients of periodicities of solar activity we will attempt to consider in the current paper.

The overview of solar activity indices defined by sunspots is presented in Section \ref{sa_ss}, the recent restoration of the new solar activity index in a grand solar cycle is described in Section  \ref{sa_eigen} while the two-millennial oscillations derived from oscillations of the solar irradiance and the baseline of eigenvectors of SBMF are described in Section \ref{sa_mil} with the discussion and conclusions to be drawn in section \ref{conc}.

\section{Spectral analysis of the averaged sunspot numbers} \label{sa_ss}
\subsection{Wavelet and Fourier spectral analysis} \label{sec:wl}
The periods of oscillations can be derived from various datasets:  averaged sunspot numbers, summary and modulus summary curves, solar irradiance, etc. using the spectral methods of Wavelet transform and Fourier analysis. Wavelet transform is a powerful tool to analyze time-series data collected by a pinpoint in the domain, to study the fluctuations in the frequency domain \citep{Mazzarella_Palumbo1989, Farge_1992, Terrence1998}. Wavelet analysis is often combined with Fourier analysis \citep{Terrence1998, Bruns2004, Aguilar2015, Thomas2022} to verify the wavelet findings. 

Wavelet analysis is an excellent tool for examining signals with time-varying frequency characteristics. Unlike Fourier analysis \citep{Terrence1998, Bruns2004, Aguilar2015, Thomas2022}, the wavelet transform gives a two-dimensional scan of the analyzed signal, while the time coordinate and frequency are independent variables. This representation allows you to explore the properties of signals simultaneously in time and frequency domains. The choice of the mother Morlet wavelet is dictated by the task to detect frequencies (or periods) and the nature of the signals under study because in the Morlet wavelet one can obtain a high-frequency resolution \citep{Farge_1992, Aguilar2015}, which is important for this task.

By considering the time series in the frequency-time space it is possible to derive dominant periods and their variations in time. Usually,  the wavelet transform revealed richer features in the high-frequency region.
For comparison in each wavelet panels we present the global wavelet spectra and the original Fourier spectra and Fourier spectra averaged by a period unit. The global wavelet spectrum is obtained by the integration of the coefficients of the wavelet transform over the time scale. The global wavelet spectra represent the distribution of temporal series over the duration-of-period axis and secure the noncontradictory estimations of power spectra of the series while reliably characterising temporal variations of the series.

To reflect the errors appearing during the wavelet analysis because of limited statistical data at the start and finish of the time series (border effects), the black dashed line marks the cone of influence (COI) defining the parts of the spectrum where these boarder effects become essential.  For this reason the results outside the COI are excluded from  further investigation. This is applicable, in particular, in calculations of the global wavelet spectrum. Global wavelet spectra are plotted by the black solid lines with COIs plotted by the black dashed lines, while the Fourier spectra are marked by indigo lines. 

The significance level of the global wavelet spectrum is marked by the dashed red lines denoting the red noise spectra within the confidence interval of $95\%$. The peaks of the global wavelet spectra above the red noise with confidence level of $95\%$ are considered to be significant.

\subsection{Solar activity index by averaged sunspot numbers}  \label{sa_index}

A system of the number of sunspots per day was introduced \citep{Wolf1877} by using the individual observations by preferred observers to compute, first, monthly averages, then yearly averages from the averages of all months, if in this year at least one observation is carried out.  The  very detailed sketches of the structure of sunspots were made by the observers with the quality not  lower in detail than with modern telescopes. 

A significant step in improving the sunspot series was made in 1998 by \citet{hoyt1998a, hoyt1998b} who published a revised sunspot series with sunspot groups (GSN) from 1610 based on the analysis of 455242 records of 463 observers. HS98 used a special “fill-in” procedure to fill sunspot numbers for days with no observations, in order to “bridge” the data gaps. 

A number of authors tried to understand the differences between the group numbers (GSN) and Wolf’s sunspot numbers (WSN)  \citep{Hoyt1994, Clette2014, velasco2021, Brehm2021}. 
The GSN series of \citet{hoyt1998b} is found to be more consistent and homogeneous with Schwabe’s data throughout the entire studied period as found by \citet{Leussu2013} while the WSN records decreased by roughly 20$\%$ around 1848 because of the change of the primary observer from Schwabe to Wolf. 
Although,  the GSN reconstruction becomes very similar to Wolf’s reconstruction before the 1.25 correction factor was applied \citep{Hoyt1994}.

\citet{Clette2014} reported a noticeable trend found and eliminated in the solar activity index derived from the observations of the Locarno Observatory, which was a reference observatory after 1980. Also \citet{Clette2014} derived the three-peak shape (so-called $\Psi$-type distribution) of the original GSN by \citet{hoyt1998a, hoyt1998b} for sunspot Cycle -1 with the peaks in 1736, 1739, and 1741.  Later a modified single peak shape for this solar cycle was suggested by a number of authors \citep[see, for example][]{usoskin2004, Vaquero2007a, Vaquero2007b, Vaquero2014} after more historical records of sunspot counts were discovered. Although, the derivation by \citet{Clette2014} indicated that the real shape of cycles in early years was not yet confirmed.

\begin{figure}
\includegraphics[scale=0.52]{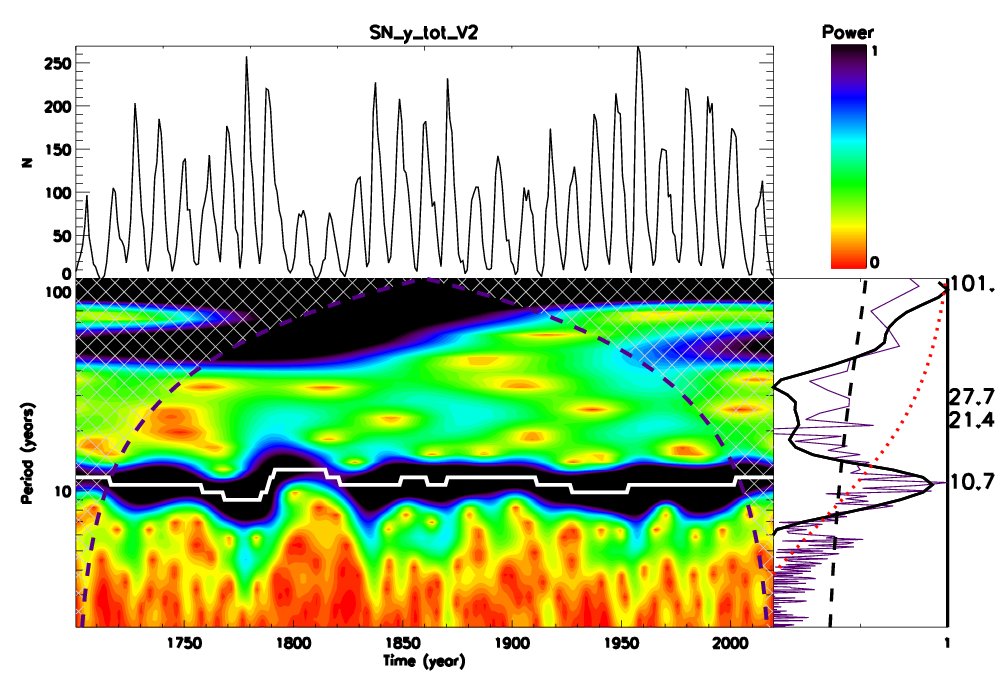}
 \caption{Top plots: Left - The averaged sunspot numbers $N$ versus time in 1700-2020 including purely defined cycles.  Right - The power bar for the wavelet spectrum. Bottom plots: Left - Wavelet spectrum with periods (Y-axis) derived from sunspot numbers, the black dashed line shows the Cone of Influence (COI) (see the text for details). Right - The global wavelet spectrum (the solid black line),  Fourier spectrum (the indigo line). The COI shown by the black dashed line and the red-noise level at 95$\%$ shown by the red dashed line. A courtesy of \citet{Zharkova2022b}. }
  \label{wl_ss2}
\end{figure}

Then an almost 400-year history of sunspot activity from 1610 to the 2000s was revised by joint efforts of researchers  \citep{Svalgaard2016, Svalgaard2017a, Svalgaard2017b}. The project used the sketches of sunspots by Christoph Scheiner, Johann Kaspar Staudacher, Heinrich Schwabe, Rudolf Wolf, and Hisako Koyama \citep{Carrasco2020, Hayakawa2020}. The authors used two backbones: the Schwabe (1794 – 1883) and the Wolfer (1841 – 1944) backbone \citep{Svalgaard2016}. 

Since July 2015, the SILSO International Data Center (Sunspot Index and Long-term Solar Observations) at the Royal Observatory of Belgium maintains a new, revised series of relative sunspot numbers SSN (Version 2.0),
which was considered to be fairly reliable since 1750 \citep{Clette2014, Leussu2013, Svalgaard2016}. This solar activity index SSN V2 will be used here for further analysis.
The monthly smoothed sunspot numbers   SSN version 2.0 \citep{SILSO} were used in this study as shown in  Fig. \ref{wl_ss2}, top plot.   

\subsection{Periods derived from the averaged sunspot numbers} \label{ss_wl}
Now we wish to present periodicities embedded into the well-used averaged sunspot numbers observed in the past 400 years as derived earlier \citep{Zharkova2022b}. 
The Morlet wavelet analysis was applied to the frequencies of averaged sunspot number reported in section \ref{sa_ss} and the result is plotted  Fig. \ref{wl_ss2}. The black solid line reflects the global wavelet spectrum, the black dashed line reflects the COI  and the indigo line represents the Fourier spectrum. The red dashed line shows the red noise level of the data within the confidence interval of $95\%$. The peaks of the global wavelet spectra above the red noise with confidence level of $95\%$ are considered to be significant.
 
The wavelet spectrum of the temporal series of averaged sunspot numbers for the whole series reveals the powerful peak at 10.7  years (corresponding to an 11-year cycle) and some smaller peaks at its double harmonics with a 21.4-year period. There are no other significant frequencies, or periods, detected besides a well-distinguished peak at 101 years, possibly, associated with Gleisberg's centennial cycle. This outcome is well expected since the sunspot data time limitations by default cannot carry any large-scale periods of higher frequencies because the dataset is rather short and contains only positive numbers of sunspots with t considering the magnetic polarities of magnetic fields in them.

This finding supports the objections expressed by the analysis of stochasticity of the averaged sunspot numbers \citep{Cameron2017} proving that  10.7 years is a real period of the oscillation of sunspot activity which can only reflect the positive variations of these sunspot numbers. In order to obtain other periods, we need to convert the data to the eigenvectors of SBMF which also include its polarity. The whole bulk of research discussed in the introduction was dedicated to the interpretation of this dataset and this period of dynamo wave generation.

\section{Solar activity index by eigen vectors of SBMF} \label{sa_eigen}
\subsection{Pair of principal components  and their summary curve} \label{ev_proxy}
 The dynamo mechanism, which governs solar activity, operated with poloidal and toroidal magnetic fields \citep{parker55}, with the first one being the solar background magnetic field (SBMF), and the second one being the magnetic field of magnetic loops in active regions, which are embedded into the solar surface, whose roots are seen as sunspots.  The interaction between these two magnetic fields defines the variations of solar activity seen through the appearance or disappearance of sunspots and active regions.
 
  However, because the SBMF is shown to be in anti-phase with the leading polarity of a magnetic field in sunspots \citep{Stix76, zharkov08} defining the locations and timing of sunspot appearances on the solar surface and their migration towards the solar equator or poles \citep{zharkov08}, one can expect that the summary curve of the SBMF should reveal defined links with the averaged sunspot numbers.  
 \begin{figure}
\includegraphics[scale=0.82]{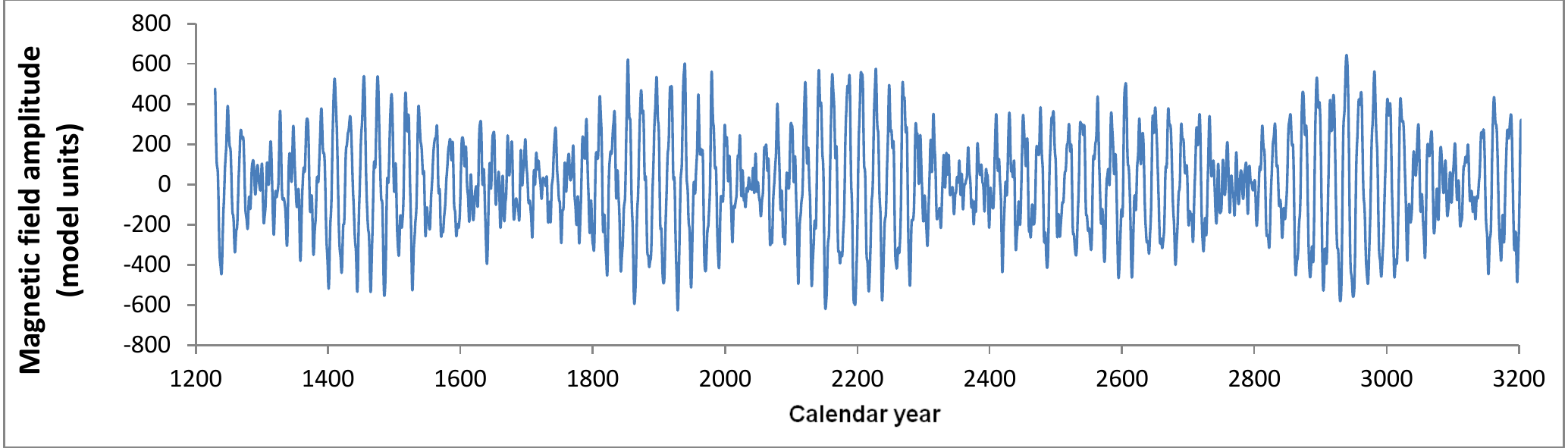}
\includegraphics[scale=0.8]{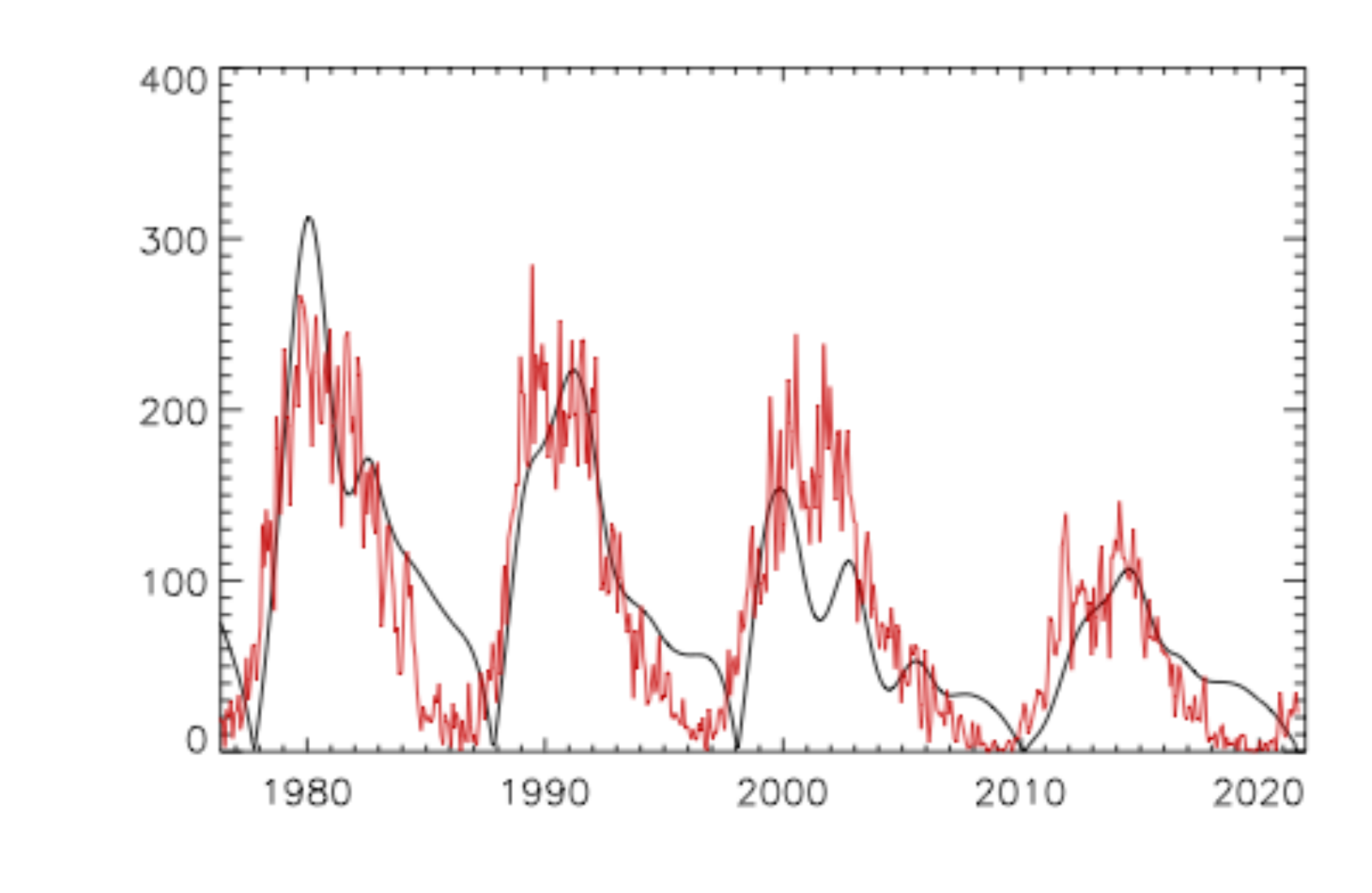}
 \caption{Top plot: The summary curve (Y-axis, arbitrary units) of two PCs calculated for 2000 years \citep[courtesy of][]{zhar15, zhar2020}. Bottom plot: the modulus summary curve (Y-axis, in arbitrary units) variations versus calendar years (X-axis) for solar cycles 21-24 (black line) compared with the averaged sunspot numbers (red line) \citep[courtesy of][]{Zharkova2022}.  }
  \label{sum}
\end{figure}   
 \citet{shepherd14, zhar15} investigated the two eigenvectors with the largest eigenvalues (39$\%$ by variance), or two principal components (PCs), or eigenvectors, of the solar background magnetic field  (SBMF) by applying the principal component analysis (PCA) to the Wilcox Solar Observatory low-resolution full disk synoptic magnetic maps for cycles 21-24. The authors identified numerous eigenvalues and eigenvectors of own magnetic waves of the Sun seen on the solar surface, which came in pairs. The four significant pairs cover the majority ($>95\%$ of the data by variance \citep{zharkova12, Zharkova2022}. 
 
The first pair, or two principal components, reflect the primary waves of the solar magnetic dynamo produced by the dipole magnetic sources \citep{zhar15}. These two waves are found traveling slightly off-phase from one hemisphere to another and their interaction defines the solar activity in each hemisphere and as a whole \citep{zharkova12}.    \citet{shepherd14, zhar15} used the symbolic regression analysis  \citep{Schmidt09} of these two magnetic waves and obtained the analytical expressions for the magnetic (dynamo) waves incorporated into the ensemble of waves present in the solar background magnetic field attributed to the poloidal field of the Sun \citep{popova13}.  These mathematical equations were used to make predictions in time by a thousand years both forwards and backwards, from the current epoch and to use them for a comparison with the magnetic waves supposedly produced by the solar dynamo acting in two layers  with slightly different meridional circulation velocities \citep{zhar15}.  

In order to bring the detected trends in the SBMF closer to the currently-used index of solar activity, the averaged sunspot numbers, we calculated the summary component of the two PCs. The modulus summary curve was found to correlate closely with averaged sunspot numbers \citep{shepherd14, zhar15}. \citet{zhar15} suggested using the summary curve of these two PCs as a new proxy of solar activity, instead of, or in addition to, the averaged sunspot numbers. This suggestion was confirmed by previous research with PCA  of the solar magnetic synoptic maps of Kitt Peak observatory obtained with much higher resolution \citep{Zharkova2022}, which PCs, in fact, reflect the magnetic fields of active regions, e.g. toroidal magnetic field, well known to have the 11-year periodicity classified via averaged sunspot numbers. While the lower resolution  SBMF from WSO used in our previous research \citep{zharkova12, zhar15, Zharkova2022} PCA detects the two PCs, or magnetic waves, of the poloidal magnetic field associated with dipole magnetic sources \citep{zhar15} and another three significant pairs associated with quadruple, sextuple and octuple magnetic sources \citep{Zharkova2022}.

Using the derived formulae, the summary curve was calculated backward to 1200  and forward to 3200 \citep{zhar15} as shown in Fig. \ref{sum}, top plot revealing very distinct variations of the cycle amplitudes in every 330-400 years, or grand solar cycles (GSCs). These grand solar cycles are separated by grand solar minima (GSMs),  when the amplitudes of 11-year cycles become very small, similar to those reported in Maunder,  Wolf and Oort and other grand solar minima \citep{Zharkova2018b, Zharkova2018a}. 

The timings of the grand solar minima are shown \citep{zhar15} to be defined by the interference of two magnetic dynamo waves generated in different layers with close but not equal frequencies defined by the different velocities of meridional circulation {so-called beating effect).  The calculation of the summary curve forward in time until 3200 has shown the further three grand solar cycles separated by the three GSMs with the first GSM  to occur right now during the cycles 25-27, or in 2020-2053  and the next one in 2375-2415 \citep{zhar15, zhar2020}. 

\subsubsection{Modulus summary curve and the sunspot index} \label{ev_ssn}

The modulus summary curve (MSC) reflecting the summary curve to the positive plane is shown in Fig. \ref{sum}, the bottom plot for cycles 21-24 \citet{Zharkova2022}.
 The temporal features of a modulus of the summary curve of these two PCs show a remarkable resemblance to the sunspot index of solar activity for cycles 21-23 \citep{shepherd14, zhar15}, or recently cycles 21-24 \citep{Zharkova2022}.   
 
  Embracing the similarity between the modulus summary curve  (MSC) and averaged sunspot numbers (SSN), the modulus summary curve can be normalized for each cycle by the averaged sunspot numbers indicated by the left Y-axis. The modulus curve, in general, follows the averaged sunspot numbers for all the cycles revealing a significant reduction of solar activity from cycle 21 (maximum about 300 sunspots), through cycle 22 (230), 23 (165)  to cycle 24 (108) \citep{Zharkova2022} that fits reasonably to the maximum numbers reported for cycles 21-24 \citep{SILSO}: 21 - 233, 22 - 213, 23 - 180, 24 - 116. The further predictions of the MSC reveals the continuing reduction of solar activity and the occurrence of grand solar minimum in cycles 25-27 \citep{zhar15}.
 
There was a remarkable resemblance between these two curves, SSN and MSC shown in Fig. \ref{sum}, bottom plot \citep{zhar15, Zharkova2022, Zharkova2022b}. This similarity occurs despite the modulus summary curve reflecting the poloidal magnetic field while sunspots - toroidal one, which allowed the authors to suggest this summary curve of PCs, or eigenvectors of SBMF, as a new solar activity proxy. The advantages of using the solar index from the summary curve instead of the averaged sunspot numbers are the ability to do long-term predictions and the presence of the extra-parameter, a leading polarity of the background magnetic field of the Sun.  
  
  Hence, on the one hand, the modulus summary curve is found to be a good proxy of the traditional solar activity index contained in the averaged sunspot numbers predicting the grand solar minimum in the the next three decades (2020-2053). This suggestion of reduced solar activity in cycles 25-27 was also supported by the recent research of the same SBMF data of WSO \citep{Kitiashvili2020, Obridko2021} and by application to the series of sunspot numbers of the singular spectral analysis (SSA) \citep{Courtillot2021} or Bayesian method \citep{velasco2021}. At the same time, the summary curve proposed by \citet{shepherd14, zhar15}  as a sum of the largest eigenvectors of SBMF, which were given mathematical description via a series of cosine functions, is shown to represent a real physical process - poloidal field dynamo waves generated from dipole magnetic sources by the solar dynamo in two layers of the solar interior \citep{zhar15}. 
 
  Therefore, the modulus summary curve proves that the eigenvectors of SBMF can be considered as a very good proxy of the traditional solar activity understandable by many observers. It should be used as a complementary solar activity index in addition to the existing one of the averaged sunspot numbers. Plus, this new index adds the additional parameter to this proxy of a dominant polarity of the solar background magnetic field for each cycle, which has the polarity opposite to the leading polarity of sunspots \citep{Stix76, zharkov08}.

 Based on the similarity of the modulus summary and sunspot curves, one can conclude that the solar activity in cycles 21-24 is systematically decreasing with cycle number \citep{zhar15, Zharkova2022} because of the shift in phase of the two magnetic waves so that their phase difference is increasing in time, approaching an anti-phase when there is a lack of any interaction between these two dynamo waves. This wave separation into the opposite phases will definitely lead to the absence of active regions, or magnetic flux tubes, whose roots appear on the solar surface as sunspots. This, in turn, can lead to an absence of noticeable activity on the solar surface, especially, in the descending and minimal phases of cycles 25-27 \citep{zhar15} that can resemble the similar features recorded during the Maunder Minimum  \citep{Eddy1976}.

\subsection{Periods of oscillation derived from the eigenvectors} \label{wl_eigen}
\subsubsection{Spectral analysis of the summary curve} \label{wl_summ}
In order to extend the frequency detections, let us move to the dataset of eigenvectors derived as the summary curve of the two largest eigenvectors of the solar background magnetic field (poloidal field) \citep{zhar15}.  We first analyzed with the wavelet and Fourier transforms the shorter curve of 1200 years (2000-3200 (see Fig. \ref{wl_short} )and then the full curve of 2000 years (1200-3200) presented in our first paper \citep{zhar15} (see Fig. \ref{wl_sc2000}).  The results are shown with the original curve plotted in the top plots, Fig. \ref{wl_short} and Fig. \ref{wl_sc2000},  the wavelet image shown in the bottom left plots and in the periods detected with the global wavelet and Fourier analysis, which are presented in the right plots.
\begin{figure}
\includegraphics[scale=0.2]{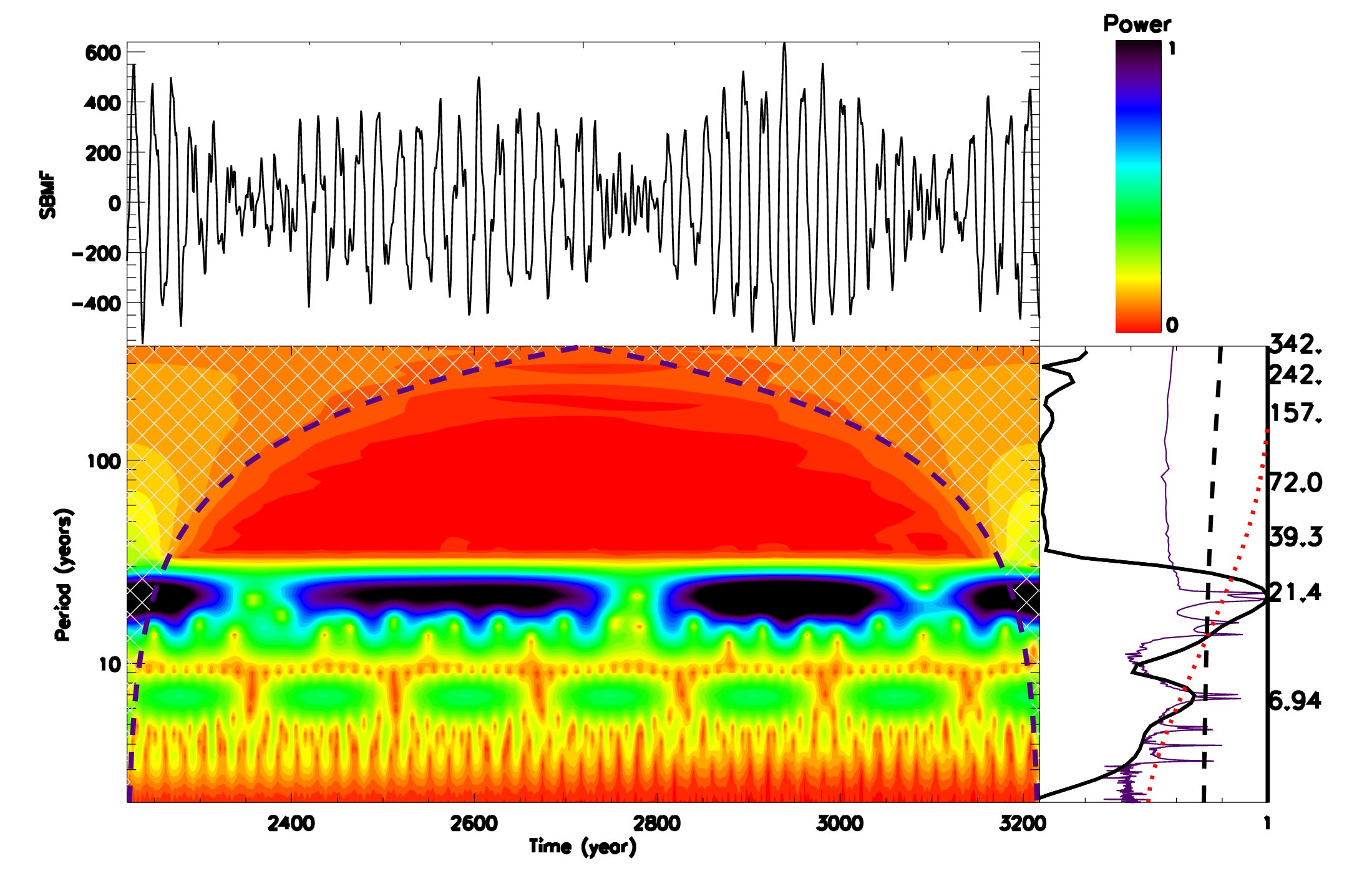}
 \caption{Left: Wavelet spectrum  (bottom plot) with a power shown by the color bar of the summary curve, E  (Y-axis, in arbitrary units) (top plot) of the two largest eigenvectors of SBMF  from Fig. \ref{sum}, the top plot for the years 2200-3200. Right: Global wavelet spectrum is plotted by the black line, the COOI indicated by black dashed line, the Fourier spectrum is marked by the indigo line and the red noise within 95$\%$ confidence level is shown by the red dashed line (see for details section \ref{sec:wl}).  The right Y-axis shows the derived periods in years. The period of 21.4 years is confidently located above the red noise level. }
  \label{wl_short}
\end{figure}
\begin{figure}
\includegraphics[scale=0.2]{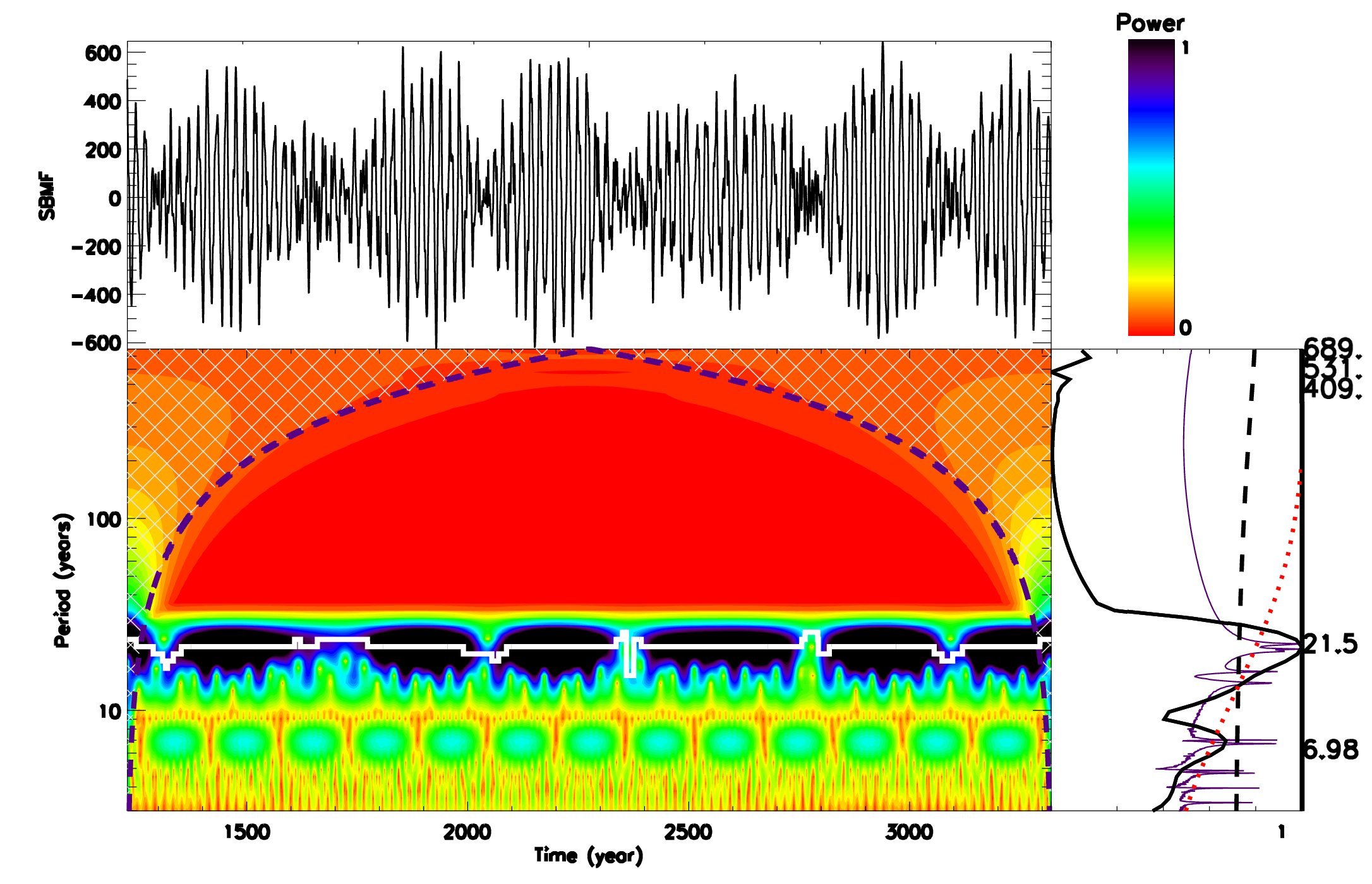}
 \caption{ Left: Wavelet spectrum  (bottom plot) with a power shown by the color bar of the summary curve, E (Y-axis, in arbitrary units) (top plot) of the two largest eigenvectors of SBMF  taken from Fig. \ref{sum}, top plot, for the years 1200-3200. Right: Global wavelet spectrum is plotted by the black solid line,  the COI - by the black dashed line, the Fourier spectrum is marked by the indigo line, and the red noise at 95$\%$ confidence level is shown by the red line (see for details section \ref{sec:wl}). The Y-axis shows the derived periods in years.}
  \label{wl_sc2000}
\end{figure}

 It can be seen that for the summary curve of eigenvectors having as positive (northern polarity) so negative (southern polarity) magnitudes, none of the methods can detect any periodic oscillations except for 22 years  (21.5 years, precisely) for the full set of the data (Fig. \ref{wl_sc2000}). The same result with the most prominent cycle being about 21.4 years is achieved on a shorter timescale of the summary curve for 2200-3200 years shown in Fig. \ref{wl_short}. There is an indication in the Fourier spectrum of the detection of a larger cycle of 342 years in the short and 409 years in the full 2000 years of data.  This indicates that, in order to detect the 11-year cycle, we need to convert the summary curve into a modulus summary curve and then repeat the analysis.

\subsubsection{Spectral analysis of the modulus summary curve (MSC)} \label{wl_msc}
 
\begin{figure}
\includegraphics[scale=0.52]{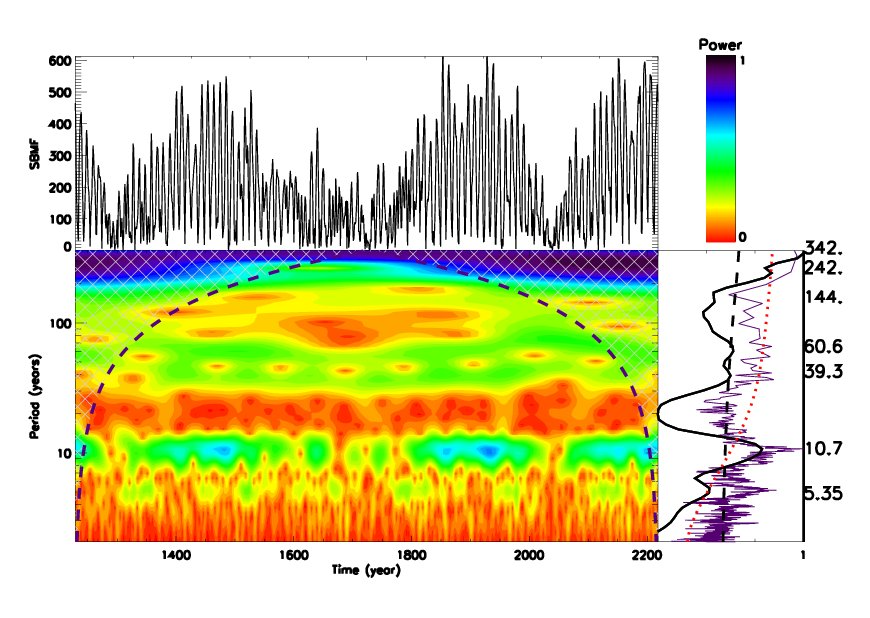}
 \caption { Left: Wavelet spectrum  (with a power shown by the color bar (bottom plot) of the modulus summary curve,  E, (Y-axis in arbitrary units) (top plot) of the two largest eigenvectors of SBMF for the years 1200-2200 (X-axis). Right: Global wavelet spectrum is plotted by the black line, the Fourier spectrum is marked by indigo line, and the Fourier spectrum averaged by a period is shown by the green line. (see for details section \ref{sec:wl}). The right Y-axis shows the derived periods in years.   A courtesy of \citet{Zharkova2022b}. }
  \label{wl-mod1}
\end{figure}

\begin{figure}
\includegraphics[scale=0.2]{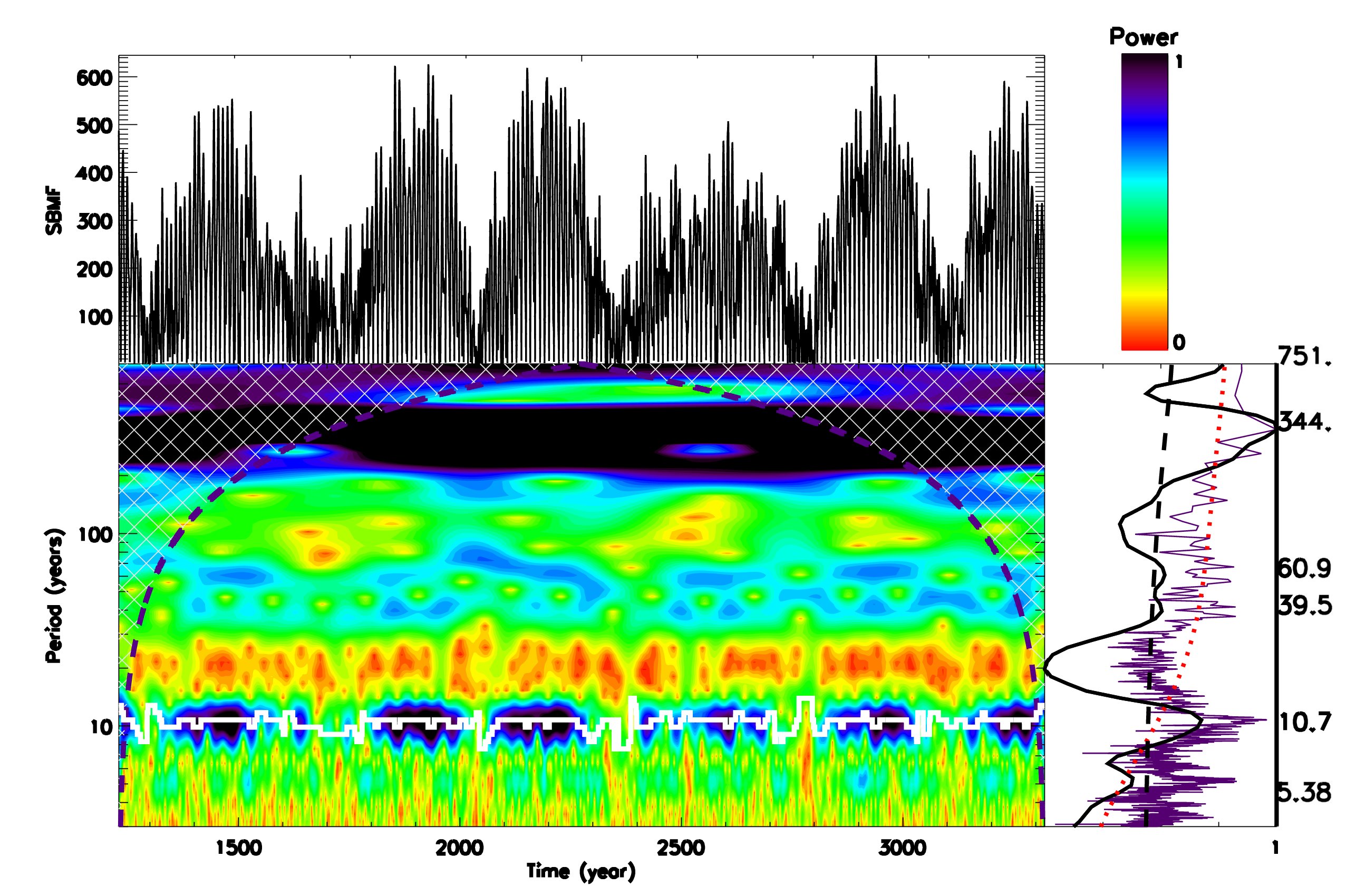}
 \caption{Left: Wavelet spectrum (with a power shown by the color bar) (bottom plot) of the modulus summary curve E  (Y-axis, in arbitrary units) (top plot) of the two largest eigenvectors of SBMF for the years 1200-2200. Global wavelet spectrum is plotted by the black  solid line, the COI - by the black dashed line, the Fourier spectrum is marked by the indigo line, and the red noise at 95$\%$ confidence levels is shown by the red line (see for details section \ref{sec:wl}). The Y-axis shows the derived periods in years. }
\label{wl-mod2}
\end{figure}
Now let us analyze with Morlet wavelet the modulus summary curves for these two summary curves of 1000 years (2200-3200)  and of 2000 years (1200-3200) taken from our paper \citep{zhar15}. The results are shown in Fig. \ref{wl-mod1} for the shorter period of 2200-3200 and in Fig. \ref{wl-mod2} with the original curve plotted in the top plot, the wavelet analysis image in the bottom left plot and frequencies detected are presented in the bottom right plot confirmed with the Fourier spectrum.

The original modulus summary curves are plotted in the top plots, the wavelet spectrum in the bottom left plot, the color bar of the wavelet powers in the top right plot, and the global wavelet spectrum (solid black line) and Fourier spectrum (solid Indigo line)  in the bottom right plot.  
 The black dashed line in the wavelet and global wavelet spectrum marks the Cone of Influence (COI) defining the parts of the spectrum where the border effects of a wavelet analysis become essential and, thus,  excluded from the further investigation. 

 The significance level of the global wavelet spectrum is marked in Fig. \ref{wl-mod1}, the right bottom image, by the dashed red lines denoting the red noise spectra within the confidence interval of $95\%$. Note that the power of the red noise grows with a reduction of frequency, e.g. with the growth of a period of oscillations. The peaks above the confidence level of $95\%$ of the global wavelet spectra (dashed lines) are considered to be significant.
 
 It can be seen that for the modulus summary curve of eigenvectors all having positive magnitudes, we detect the 11-year cycles (10.7$\pm$1.4 years, precisely), similar to that derived from the sunspots. In addition, there are smaller maxima appearing on the harmonics of the 11-year cycle and a very strong maximum at 342 years for the shorted MSC and 344 years for the longer MSC of 2000 years.  This corresponds to the averaged period of the grand solar cycle detected by \citet{zhar15} from the pair of eigenvectors derived from the SMBF.
 
 \begin{figure}
\includegraphics[scale=0.33]{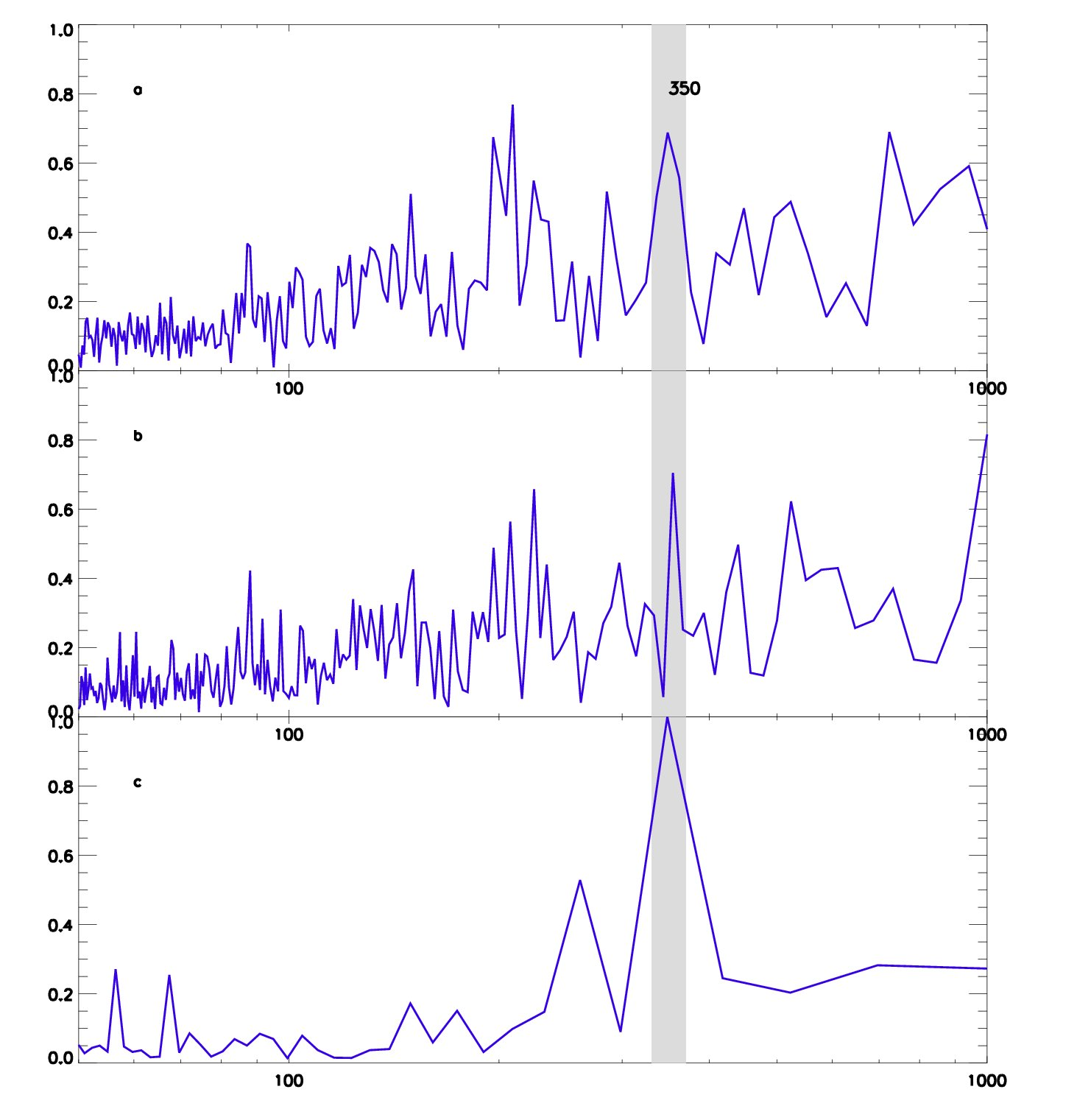}
 \caption{Comparison of Fourier spectra (Y-axis) of solar irradiance variations over the detected periods (X-axis) observed during the Holocene by \citet{Steinhilber12} (top plot), \citet{vieira2011} (middle plot) and derived from the modulus summary curve of SBMF  (bottom plot) obtained from  Fig. \ref{sum}, top plot, \citep{zhar15}.}
  \label{fourier}
\end{figure}
 
A similar period of oscillation of 350 years, was also found from Fourier spectra of solar irradiance observed by two different observers \citep{Steinhilber12, vieira2011} (see Fig. \ref{fourier}), it corresponds rather closely to the period of a grand solar cycle (GSC) found from the summary curve of eigenvectors of SBMF \citep{zhar15}, which are shown to be generated by the interference of magnetic waves produced by dipole magnetic sources in two (inner and outer) layers of the solar interior. 

 \subsection{Modelling solar activity with the double solar dynamo model} \label{sec:d-model}

For the interpretation of the magnetic waves derived from the eigenvectors of SBMF, the Interface Dynamo models were considered for two layers at different depths, where the dynamo waves are generated by several (dipole \citep{zhar15}or/and quadruple \citep{Popova18}) magnetic sources. These models took into account meridional flows in two layers located at different depths (one close to the bottom of the convective zone and the other close to the solar surface \citep{zhar15}. The flows are assumed to have different velocities of meridional flows as shown in observations \citep[see, for example,][]{Zhao2013} that consequently, leads to slightly different periods of the waves generated at different depths.

In the present work, we can use a low-mode approach for $\alpha-\Omega$ dynamo with meridional circulation in two layers. For modelling a complex behavior of the magnetic field the low-mode approach was proposed and developed by a number of authors \citep{Ruzmaikin1981, Kitiashvili2008, Kitiashvili2009, Popova2013, Popova18}  The main idea consists of the procedure where the system of differential equations, describing the magnetic field at the beginning of action is replaced by a dynamical system of not very high order equations selected in a certain way. In this case, we assume that one dynamo source is subsurface and another source is located deeply in the solar convection zone. The parameters that describe the magnetic field generation in each layer can be different, which can result in the simultaneous existence of two magnetic activity cycles with different periods. 

The Parker dynamic equations \citep{parker93}, describing the generation and evolution of the solar magnetic field in a two-layer medium, are obtained from the system of electrodynamic equations for mean fields \citep{Krause1980} on the assumption that the dynamo wave propagates in a thin spherical shell. In this case, the magnetic field is averaged along the radius within a certain spherical shell and the terms describing the curvature effects near the pole are deleted. In addition, in this approximation, we assume that the magnetic field is generated independently in either layer. \citet{Popova2008} indicated how meridional flows can be taken into account in the proposed dynamo model. 
As a result, the equations take the form: 

\begin{equation}
{{\partial  A_{i}} \over {\partial t}} = R_{\alpha i}\alpha _{i}B_{i} + \beta {{\partial^2  A_{i}} \over {\partial
\theta ^2}}- V_{i}{{\partial A_{i}} \over {\partial
\theta }},
 \label{park}
\end{equation}
\begin{equation}
{{\partial B_{i}} \over {\partial t}} =
  R_{\omega_{i}} \sin \theta {{\partial  A_{i}} \over {\partial \theta }} +
\beta {{\partial^2 B_{i}} \over {\partial \theta^2}}- {{\partial V_{i} B_{i}} \over {\partial
\theta }}. \label{parc}
\end{equation}

Here, $B$ is the toroidal magnetic field, $A$ is proportional
to the toroidal component of the vector potential,
which determines the poloidal magnetic field $B_{P}=- {{\partial  A} \over {\partial \theta }}$, and
$\theta$ is the latitude measured from the pole. $V_{i}$ is the meridional circulation (in terms of one
degree per diffusion unit of time). Subscript $i$ means the
layer number (1 or 2). The factor
$\sin\theta$ describes the decrease in the length of a line of
latitude near the pole. The second equation neglects
the small contribution of the $\alpha$-effect, i.e. we use the so-called $\alpha\omega$ approximation.
Curvature effects are absent in diffusion terms. It is assumed that the radial
gradient of the angular velocity does not vary with $\theta$.
In the Eq.~\ref{park}-\ref{parc} the parameters $R_{\alpha}$ and $R_{\omega}$
describe intensity of the $\alpha$-effect and the differential
rotation, respectively, $\beta$ is the coefficient of turbulent diffusion. We used a simple scheme
for the stabilization of the magnetic
field growth, namely, the algebraic quenching of the helicity.
This scheme assumes that $\alpha= \alpha_0(\theta)/(1 + \xi^2B^2)\approx \alpha_0(\theta)(1- \xi^2B^2)$, where $\alpha_0(\theta)=\cos\theta$ is the helicity in the unmagnetised medium and $B_0=\xi^{-1}$ is the magnetic
field for which the $\alpha$-effect is considerably suppressed.

Proceeding from simple considerations, we can make certain assumptions concerning the circulation type ($V_{i}$). Since we consider a one-dimensional problem for either layer, we should consequently present the meridional circulation as a function only dependent on $\theta$. If we assume that matter moving toward the poles leaves the layer where the dynamo operates, the meridional circulation vanishes at the poles and is maximal at mid-latitudes. We use $V_{i} =v_{i}\sin\theta$ as such a meridional circulation. To describe the poleward spread of matter in the upper layer and its return to the lower layer, it is necessary to specify the opposite signs to the meridional circulation in different layers.  It was found that from helioseismological observations meridional circulation has two cells  \citep{Zhao2013}.

The basic idea of the low-mode approach is that the mean-field dynamo equations are projected
onto a minimum set of several first eigenfunctions of the problem describing the decay of magnetic
fields without any generation sources. In this case, it is necessary to choose the minimum set of functions in such a way that the solution, which is a set of several first time-dependent Fourier coefficients taking into account the generation sources, will describe the general behaviour of the magnetic field of a given object and will not describe this field with any lower set of functions. Substituting a chosen set of the components of a magnetic field into the dynamo equations one can obtain a dynamical system of equations containing the selected modes.

To study the behaviour of harmonics in our problem, consider the case $n=10$, for~which the dynamo number of the Sun is in the range of applicability of the method. We also took into account a larger number of harmonics and found that their contribution to the solution is negligible. The toroidal field $B$ and the vector potential $A$ can be represented as:

\begin{equation}
B(\theta,t)=\sum_{n=1}^{10} b_n(t)\sin{(2n\theta)}
\end{equation}

\begin{equation}
A(\theta,t)=\sum_{n=1}^{10} a_n(t)\sin{((2n+1)\theta)}
\end{equation}

Substituting the chosen type of solution into the system (1) and (2), we obtain a dynamic system of 20 equations for unknown functions $a_n (t)$ and $b_n (t)$. Since this system turns out to be too cumbersome, we do not present it here.


We have studied the behavior of waves for dynamo numbers corresponding to solar plasma to be about $-10000$. Calculations showed that the dependence of the cycle duration on the meridional circulation is almost linear, taking into account errors in normalization. The temporal distributions of poloidal and toroidal magnetic fields produced by dipole magnetic sources are shown in Fig. 6 of the paper \citet{zhar15}. Here we present a comparison fit of the poloidal magnetic field derived from the dynamo model (Fig. \ref{dyn_mod}, bottom plot) to the summary curve derived from the eigenvectors of SBMF (Fig. \ref{dyn_mod}, top plot).  It can be noted that the simulated poloidal field variations fit rather closely the variations of amplitudes and cycle numbers in grand solar cycles derived from the eigen vectors as shown  in Fig. \ref{sum}, top plot.

The interaction of the two dynamo waves generated at different depths of the solar interior, which have a phase shift increasing with time, leads to a beating effect in the amplitude with the overall period of 330–370 years, or the grand cycles, superimposed on a standard 22-year cycle \citep{zhar15}. These grand cycles are separated by Grand Solar Minima (GSMs) when the phase difference between the waves approaches nearly half of the period. Since the time in the dynamo models is normalised to the diffusion time, which depends on the coefficient of turbulent diffusion $\beta$ in the medium, the meridional circulation velocities required to satisfy the observed grand solar cycles with the values of individual solar cycles (from 9 to 13 years) can be approximated by 9 and 10 m/s velocities in each layer for  the cycle period of 11 years, in order to achieve the required beat of 340-380 years.

\begin{figure}
\includegraphics[width=16.5  cm]{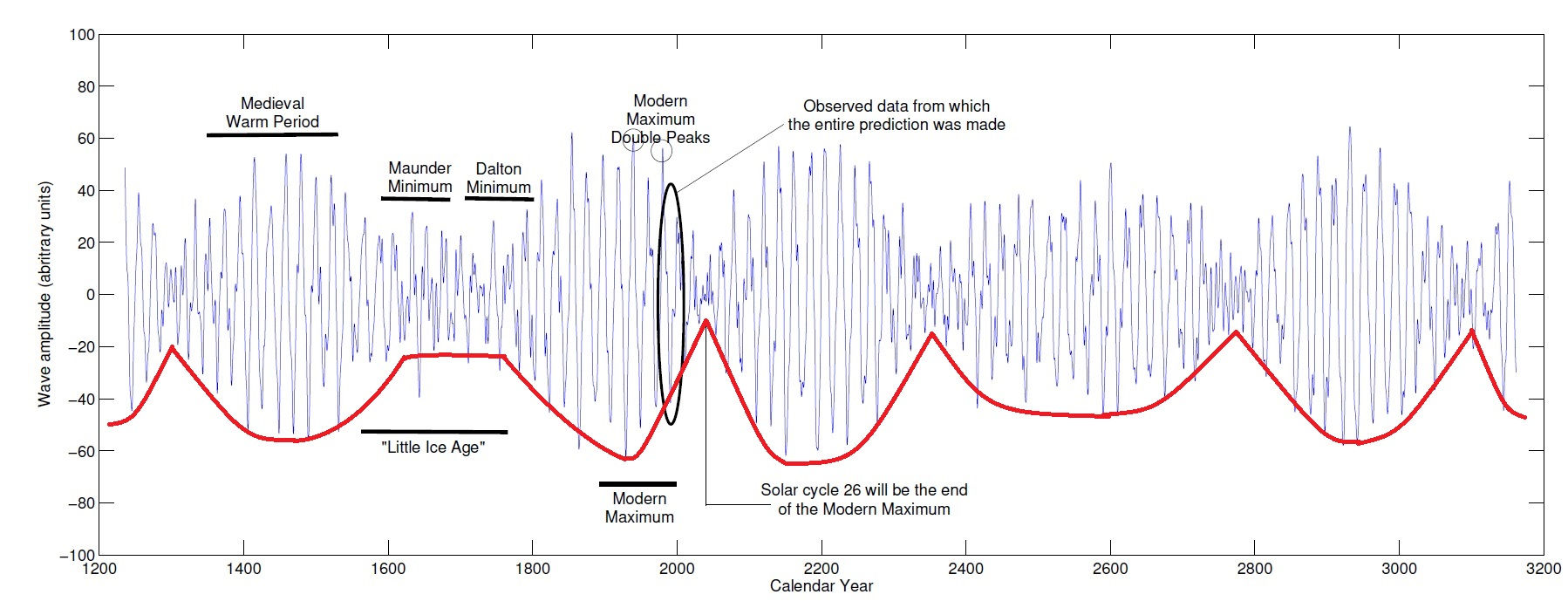}
\includegraphics[width=16.5  cm]{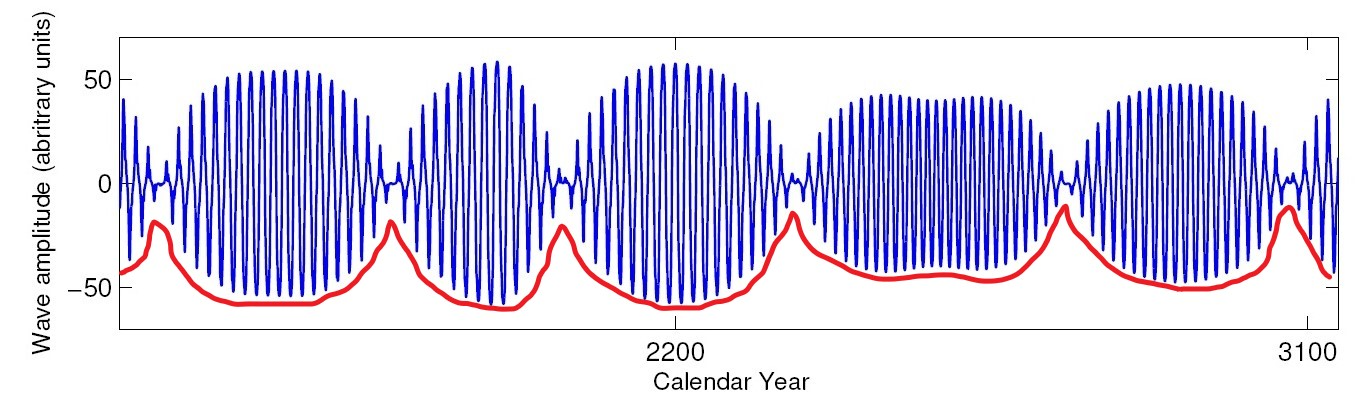}
\caption{Top plot: The predicted summary wave (the sum of two principal components) (Y-axis, arbitrary units)  calculated versus time (X-axis)  from 1200 to 3200 years from the ‘historical’ period (cycles 21–23) marked with a black oval (a courtesy of \citet{zhar15}). The historical maxima and minima of solar activity in the past are marked by horizontal brackets.  Bottom plot: Variations of the summary poloidal magnetic fields (Y-axis, arbitrary unis) simulated for 2000 years with the two layers $\alpha\Omega$-dynamo model for dipole magnetic sources with the parameters derived from the two PCs. One arbitrary unit approximately corresponds to 1–1.5 Gauss. The red lines show the envelope curves reflecting the shape of the beating effect in amplitudes of the two-wave interference.}
\label{dyn_mod}
\end{figure}  
 
 A grand solar cycle with the minima occurring every 330-380 years is likely to be a product of the beating effect of the two waves, which are generated in two different layers with slightly different meridional circulation velocities. Calculations have shown \citep{zhar15}  that for a 22 years cycle of solar activity the beating effect between the waves generated in two layers can naturally occur if the meridional circulation velocities in these two layers are about 9 and 10 m/s, respectively.  The beating effect between these waves with close periods of around 22 years can explain the variations of high-frequency amplitudes and the period of the low-frequency envelope wave with the period of 330-380 years in the resulting grand cycles seen in the observational curve (Fig. \ref{sum}, top plot). 

The beauty of this low-mode double dynamo model shown here \citep{zhar15, Popova18} is that it fits the magnetic field waves derived from the eigenvectors of magnetic waves generated by dipole magnetic sources. Having two layers allows to reconcile the Interface Dynamo models valid in the bottom layer with turbulent dynamo models working in the shallow layer under the surface without changing dramatically the parameters of the solar plasma of both layers of the solar interior (dynamo numbers, $\alpha$- and $\Omega$ parameters). This model allows us to obtain an elegant solution of dynamo equations for a very reasonable difference in the velocities of meridional circulation in these two layers not exceeding 1 m/s.

Furthermore,  \citet{Popova18} added the effect of additional magnetic waves generated by quadruple magnetic sources in these layers that allowed them to simulate the centennial (Gleissberg) cycle.   Using the dynamo model with meridional circulation and selecting the directions of circulation for quadruple waves, the authors estimate the parameters of quadruple waves best fitting the observations in the past grand cycle.

The comparison shows that the quadruple wave has to be generated in the inner layer of the solar convective zone, in order to provide the additional minima observed in the 19th and 20th centuries, thus, naturally accounting for the Gleissberg's centennial cycle. The resulting curve comprised from the summary curve for dipole magnetic sources and the simulated dynamo waves for quadruple sources reveals much closer correspondence to the averaged sunspot numbers derived in the current grand cycle. By considering the interference of two dipole and one quadruple waves, \citet{Popova18} reproduces the revised summary curve for the last 400 years which is able to account for the additional minima of solar activity, which occurred at the beginning of the 19th (Dalton minimum) and 20th centuries.  

Another period not discussed here is the quasi-biennial period of about 2-2.7 years reported from different magnetic field observations \citep{Benev1995, zharkov08}. The appearance of the quasi-biennial cycles in the observations was shown to be reproduced with a contribution of the transport dynamo, in which the meridional circulation participates in the generation of the magnetic field, transfers it, and acts to shorten the cycle periods \citep{Popova2013}.

\begin{figure*}
\includegraphics[scale=0.68]{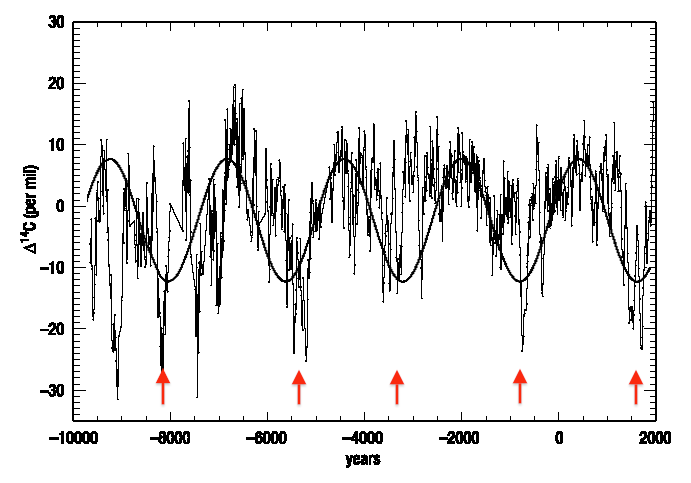}
\includegraphics[scale=0.46]{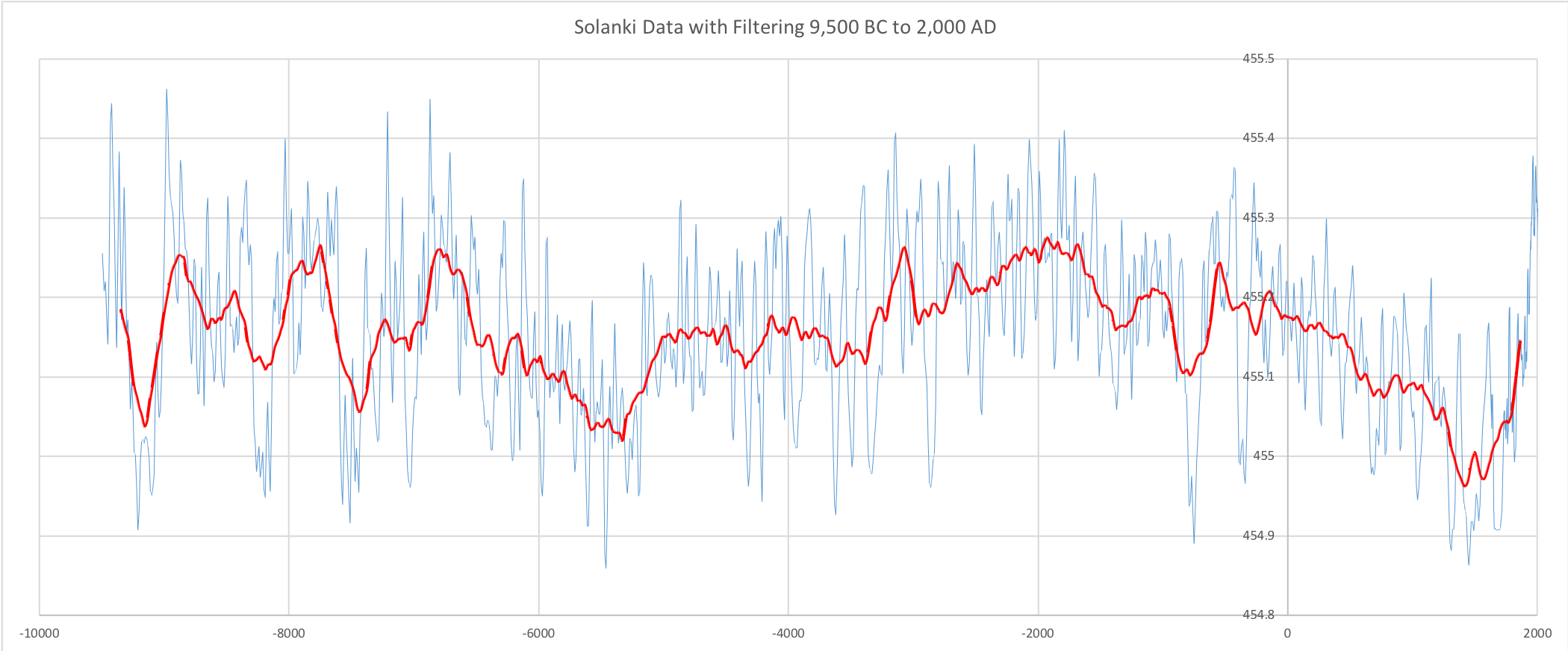}
\caption{ {\it Top plot}: the oscillations of the carbon $^{14}$C isotope abundances used by \citet{Reimer09} for solar irradiance dating in the IntCal09 data, which also reveal the Hallstatt cycle period (about 2200 years).  {\it Bottom plot}: the unnormalised solar irradiance (Y-axis) recovered for the holocene \citep{Steinhilber09, vieira2011} versus calendar yeaars (X-axis), which demonstrates  weak oscillations in the filtered (red) line with a period of about 2200-2300 years, similar to those  reported in our unfairly retracted paper \citep{Zharkova2019} and confirmed later with the real Sun-Earth ephemeris \citep{Zharkova2021}. }
\label{periods}
\end{figure*} 
 
\begin{figure}
\includegraphics[scale=0.28]{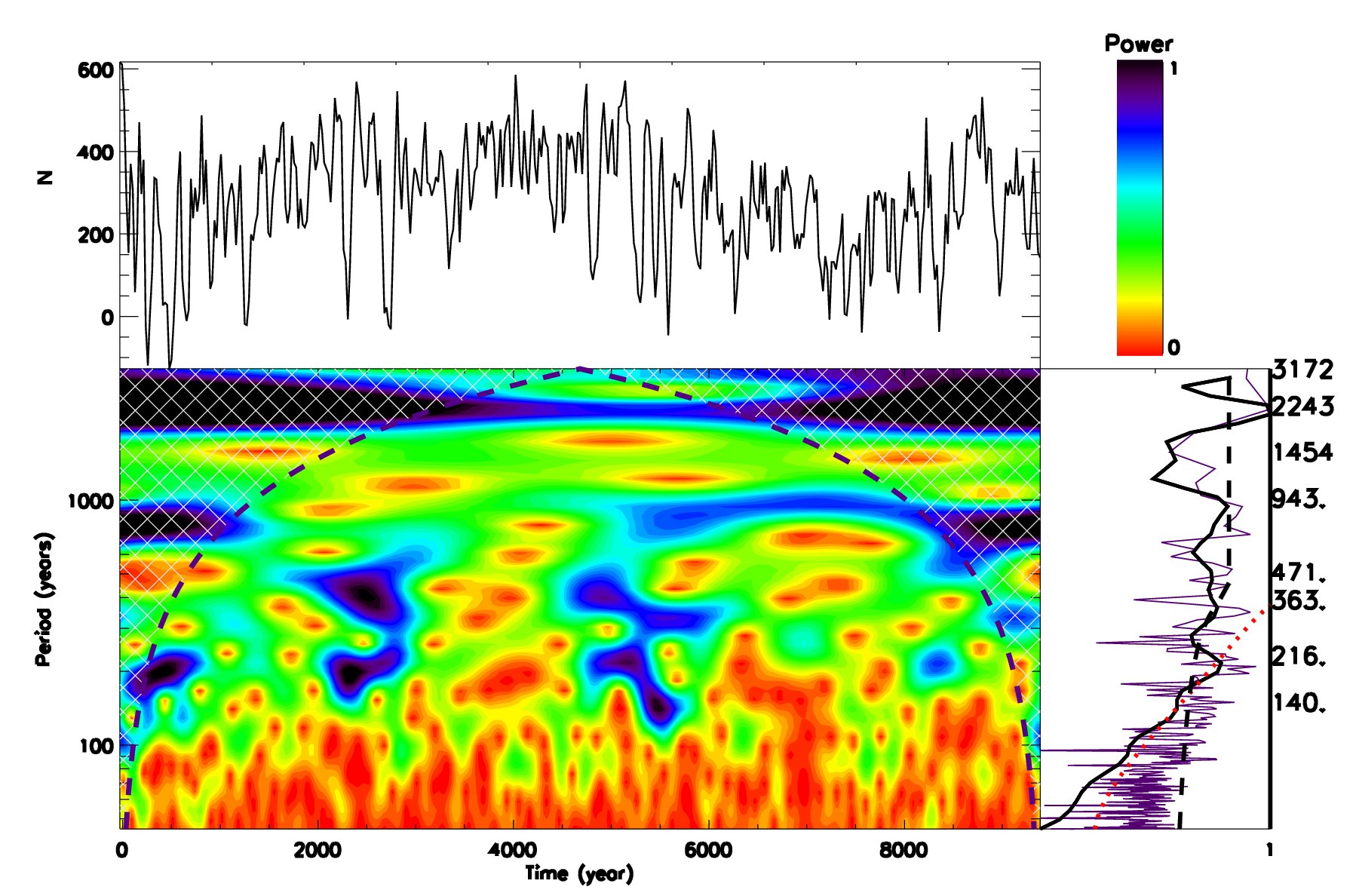}
 \caption{Left: Wavelet spectrum (bottom plot)  with a power shown by the colour bar of the total solar irradiance, N, derived for the Holocene by \citet{Steinhilber12} (top plot). Right: Global wavelet spectrum is plotted by the black solid line, the COI by the black dashed line,  Fourier spectrum is marked by thee indigo line and the red noise at 95$\%$ confidence level is shown by the red line (see for details section \ref{sec:wl}). The Y-axis shows the derived periods in years. }
  \label{stein}
\end{figure}
  
 \begin{figure}
 \includegraphics[scale=0.29]{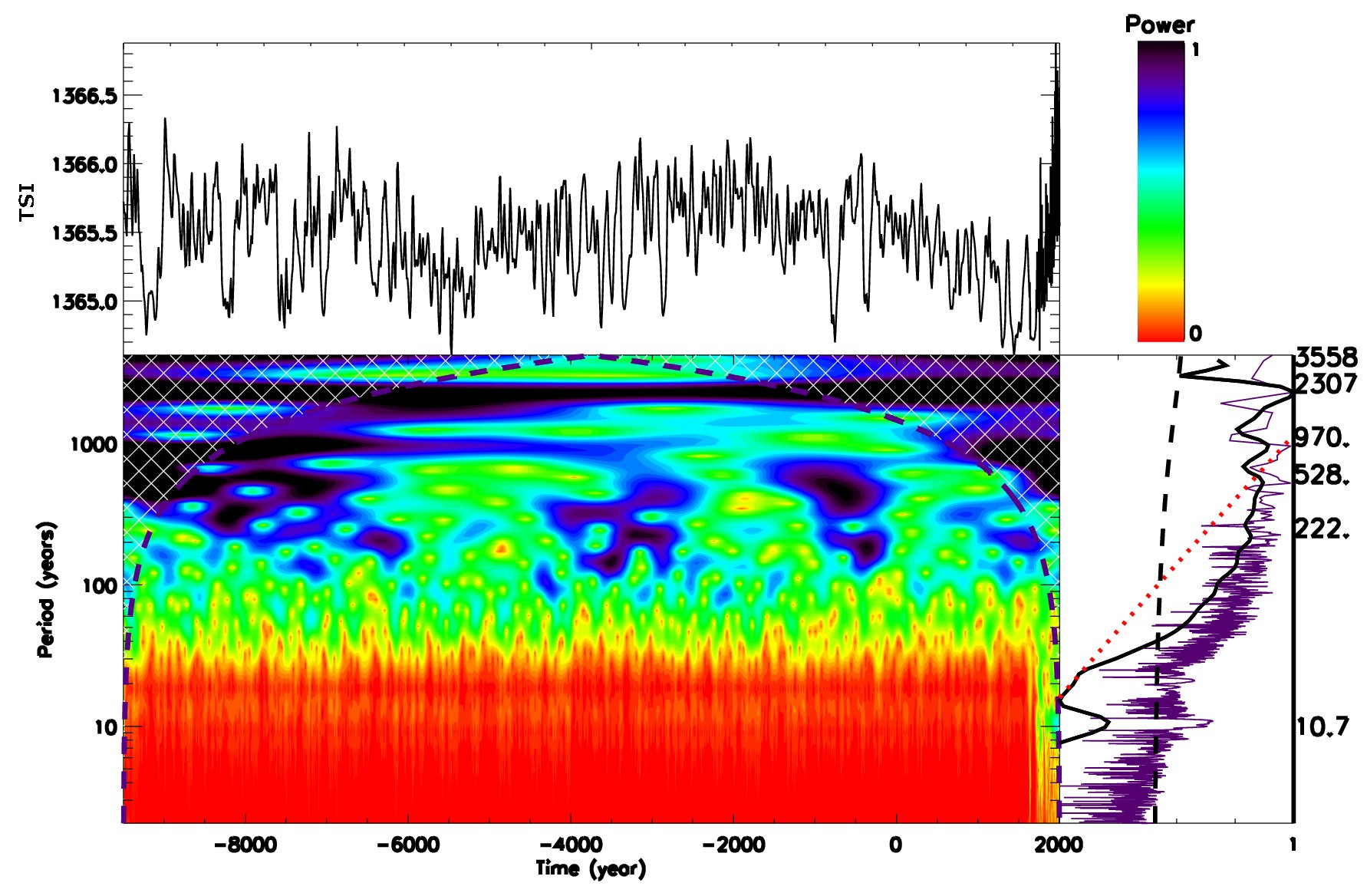}
 \caption{Left: Wavelet spectrum (bottom plot)  with a power shown by the color bar of the total solar irradiance, N, derived for the Holocene by \citet{vieira2011} (top plot). Right: Global wavelet spectrum is plotted by the black solid line,  the COI by the black dashed line, Fourier spectrum is marked by the indigo line and the red noise at 95$\%$ confidence level  is shown by the red line (see for details section \ref{sec:wl}). The Y-axis shows the derived periods in years. }
  \label{vieira}
\end{figure}

 \section{Two-millennial periods in variations of the solar irradiance and SBMF baseline}\label{sa_mil}
\subsection{Periods detected in oscillations of solar irradiance}
There is also a two-millennial period recorded in the variations of solar radiation recorded from the terrestrial biomass with the abundances of isotopes of carbon $^{14}$C and beryllium $^{10}$Be \citep{vieira2011, vieira2011Cat, Steinhilber12}.  The variations over the past 12 000 years of the carbon $^{14}C$ isotope abundance are plotted in Fig.\ref{periods} (top plot) \citep{Reimer09} and the derived solar irradiance oscillations \citep{vieira2011, vieira2011Cat}  are shown in  Fig.\ref{periods} (bottom plot). 

Let us now apply the Morlet wavelet and Fourier analysis to the solar irradiance data. 
    In Fig.\ref{stein} we present 10 000 years of the solar irradiance (top left plot) gathered from the abundances of carbon 14 isotope in the terrestrial biomass derived by \citet{Steinhilber12}, the Morlet wavelet in the bottom left plot and the Fourier spectrum derived from this solar irradiance curve. It can be seen that the wavelet transform provides a rather confident wide peak at about 2000-2300 years (the wide navy blue strip in the wavelet  spectrum top) \citep{Steinhilber12} coinciding with a sharp peak at 2243 years given by the global wavelet and Fourier spectra. Although red noise errors in the wavelet spectra become very high as shown by the reed dashed line. The Fourier spectrum thought detects this two millennium peak together with some smaller peaks between 216, 363, and 471 years, which probably can account for the grand solar cycles reported in section \ref{sa_eigen}.
    
 It should be noted that the amount of $^{14}$C and $^{10}$Be on Earth depend not only on solar activity but on many terrestrial reasons including the variations in intensity of the Earth’s magnetic field \citep[e.g.][]{usoskin2016, usoskin2021}. However, the evolution of the geometry of this field in the past is not known and cannot be confidently included  \citep{Lockwood2020}.  This definitely can increase errors for any periods defined with wavelet spectral analysis, and in particular, the exact number for the two-millennium period found in the wavelet spectrum. The validity of this two-millennial period  is confirmed by the COI and the red noise errors at 95$\%$ confidence level as shown in Fig.\ref{stein}, although the exact numbers of this period can vary. This is why the further verification of detected  wavelet periods was carried out  with Fourier analysis.

 The wavelet analysis was applied to the observations of solar irradiance observed in from 9000 BC to 2000 AD by  \citet{vieira2011} and the result of wavelet and Fourier spectra are shown in Fig. \ref{vieira}.  It can be seen that the wavelet transform provides a wide peak at about 2100-2400 years while the global wavelet and Fourier spectra give the sharp peak at 2307 years. The red noise level for this dataset is much lower  than that of \citet{Steinhilber12} being below 95$\%$ confidence level. The Fourier spectrum also managed to detect a standard 11-year cycle peak at 10.7 years and some peaks associated with the grand solar cycle between  222, 374, and 528 years.  This baseline oscillation period of 2100 years is very close to the periods of 2200-2400  years called Hallstatt cycle reported from the other observations of the Sun and planets \citep{Fairbridge1987, Steinhilber09, Steinhilber12, obridko2014, scafetta2014, usoskin2016}. 
 
 The full Fourier spectra of the solar irradiance curves by \citet{Steinhilber12}  and by \citet{vieira2011} reveal a well-detectable peak at 350 years shown in Fig. \ref{fourier} (top plot).  The nature of a grand solar cycle can be assigned to the beating effects of two dynamo waves formed in two different (inner and outer) layers of the solar interior by dipole magnetic sources as discussed in section \ref{sa_eigen}. The proposed nature of the two-millennia (Hallstatt) cycles detected in solar irradiance is discussed in the sections below. 
 
 There was another dataset  of the  oscillations of solar irradiance derived from the terrestrial biomass, whose analysis with the wavelet transform and singular spectral analysis (SSA) produced the two-millennial period of 2400 years \citep{usoskin2016}. This data  could not be accessed, so we evaluated the published plots. The evaluation has found the following problems:
 
 \begin{enumerate}
\item  The wavelet spectrum shown in their Fig.7 reveals very patched wavelet powers that is clearly seen in the coloured patterns unlike the  other two TSI dataset analysed in the current paper (see Figs. \ref{stein} and \ref{vieira}). The two millennial periods in the two datasets presented  by navy blue patterns in our wavelets  are smooth and narrow while the terracotta patterns presented by \citet{usoskin2016} are extremely wide and have a break between -4000 and -3000 years. 
\item The mean period derived from the wavelet power in \citet{usoskin2016} looks more likely to range between 2048 and 2500 years. However, without a global wavelet spectrum it is difficult to define more precisely the period magnitude. 
\item The authors do not provide the Cone of Influence (COI) of their data. 
\item There is no estimation of the red noise of the dataset. 
\item Given the very patchy wavelet powers shown in Fig. 7 \citep{usoskin2016}, the red noise of this particular dataset could be substantial. This noise would significantly affect the period derived  from it with the wavelet and SSA  methods producing the number of 2400 years versus to the numbers 2243  years \citep{Steinhilber12}, 2305  years \citep{vieira2011}. 
\end{enumerate}
Which of the periods is more precise,  it is rather difficult to say, given the complexity of building these databases from the terrestrial biomass \citep{Brehm2021} that would require further investigation.  
 
 \subsubsection{Looking for reasons: oscillations of the baseline of SBMF}
 The large two-millennial period of oscillation of solar irradiance, Halllstatt's cycle, has been also detected from the summary curve of the eigenvectors \citep{Zharkova2019, Zharkova2021}. In Fig.\ref{baseline} we present 12 000 years of the summary curve (between 10 thousand years BC and  2020 years AD.  In addition to the grand cycles of solar activity of $~$350-400 years, there are also indications of the clustered five GSCs, which are marked by the vertical lines separating larger cycles, or super-grand cycles of about two thousand years (Fig. \ref{baseline} top plot).   By comparing the semi-similar features (between the vertical lines) of the repeated six grand cycles with a total duration of about 2100 years, one can see a striking similarity of the shapes of these grand cycles in each section. It can be also noted from the irradiance curve seen in both plots of Fig. \ref{stein} that the new two-millennial cycle was likely to start 350-370 years ago, during the Maunder minimum \citep{Zharkova2019, Zharkova2021}, and the solar irradiance is currently increasing.

In order to understand the nature of these super-grand oscillations of TSI  and to derive the exact frequency/ period of this super-grand cycle, let us filter out large oscillations of 11/22 year solar cycles in the summary curve of two eigenvectors of SBMF by deriving the residual of the 22-year cycle amplitude, or a baseline magnetic field for the 22-year cycle. Theoretically, this should be a zero-line, a straight line with a zero magnitude of a magnetic field, while, in reality, the derived baseline shows the oscillations about this zero-line plotted by a dark blue curve in Fig.\ref{baseline} (bottom plot) over-plotted on the summary curve itself (light blue curve) taken from the summary curve calculated backward by 20 thousand years. The left Y-axis gives the range of variations of the baseline curve (-20,20) while the right side Y-axis produces the range for the summary curve (-500,500). 
 
It is evident that the dark blue line in Fig.\ref{baseline} (bottom plot) shows much ($\approx$25 times) smaller oscillations of the baseline of the magnetic field with a period of $T_{SG}=2000 \pm 95$ years, which is incorporated into the magnetic field measurements of the summary curve (light blue curve). The baseline oscillations show a very stable period occurring during the whole duration of the simulations of 120 thousand years, for which the summary curve was calculated. This means that this oscillation of the baseline magnetic field on a millennial timescale has to be induced by a rather stable process either inside or outside the Sun.  

This baseline oscillation period of 2100 years is very close to the 2200-2400  years period called Hallstatt's cycle reported from other observations of the total solar irradiance and planets \citep{Fairbridge1987, Steinhilber09, Steinhilber12, obridko2014, scafetta2014, usoskin2016}. 
\begin{figure*}
\includegraphics[scale=0.66]{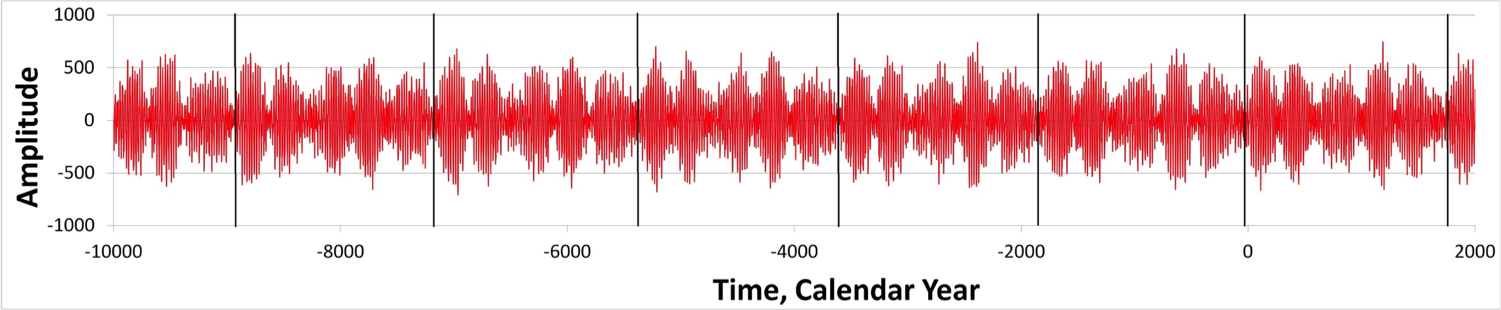}
\includegraphics[scale=0.45]{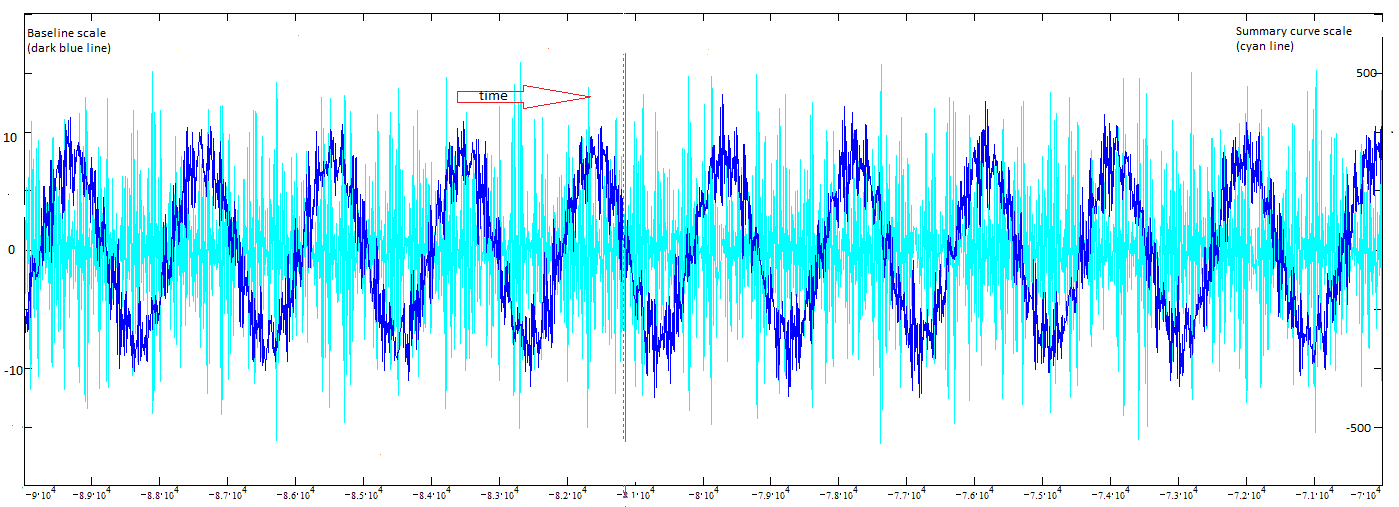}
\caption{ {\it Top plot}:  variations of the summary curve of two magnetic field waves, or PCs (Y-axis, arbitrary units), calculated backward twelve thousand years from the current time. The vertical lines define similar patterns in five grand cycles repeated every 2000-2100 years (a super-grand Hallstatt's cycle). {\it Bottom plot}:  the variations of the summary curve (cyan line) (left Y-axis, arbitrary units)  calculated backward from 0 to 20K years over-plotted by the oscillations of a magnetic field baseline, or its zero line (dark blue line) (right Y-axis, arbitrary units) with a period of about 2000$\pm$95 years. The baseline oscillations are obtained with an averaging running filter of 22 years from the summary curve suppressing large-scale 11 year cycle oscillations.  A courtesy of \citet{Zharkova2019, Zharkova2021}.} 
\label{baseline}
\end{figure*}

\begin{figure*}
\includegraphics[scale=0.73]{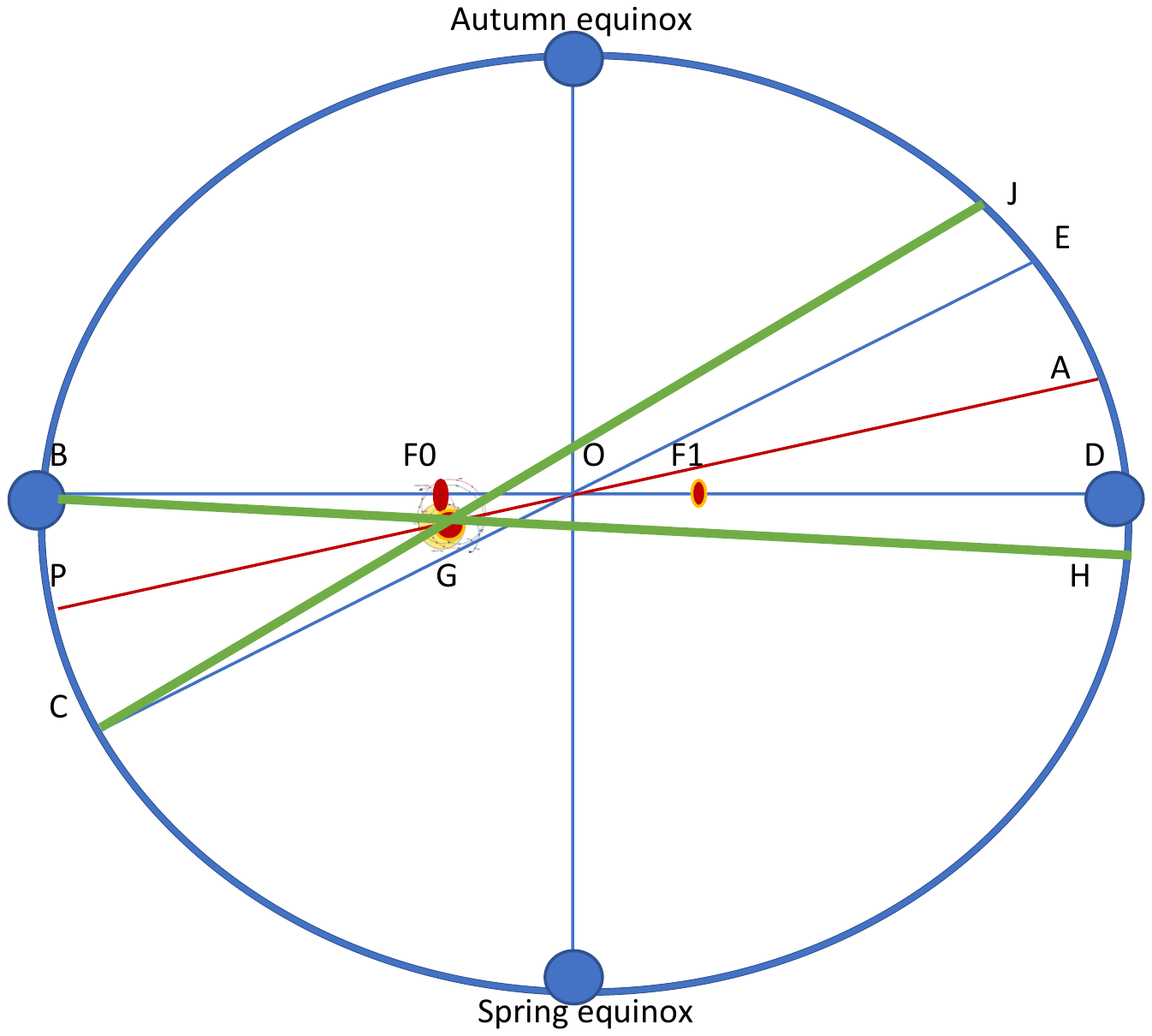}
\caption{ Orbital motion of the Earth (blue curve) about the barycentre F0 of the Solar System and the position G of the Sun defined by SIM.  See the text in section \ref{sec:dist_ephem} for details. } 
\label{e_orbit}
\end{figure*}
\subsection{Ephemeris of the Sun-Earth distances in 600-2600} \label{sec:dist_ephem}
Let us now explore the daily Sun-Earth (S-E) distances over the two millennia (600-2600) derived from the ephemeris of VASOP87 - Variations Seculaires des Orbites Planetaires \citep{Bretagnon1988} $http://neoprogrammics.com/vsop87/planetary_distance_tables/.$ Note that the VASOP87 data up to 6 digits after the decimal point coincide with the widely used JPL ephemeris \citep{Folkner2014}.  The motion of the Earth about the barycentre of the Solar System and the Sun is demonstrated in Fig. \ref{e_orbit} showing the elliptic orbit with two focuses F0 and F1 and the real position G of the Sun closer to the spring equinox imposed by the current position SIM \citep{cionco2018, perminov2018}. 

In this setting, the real Sun-Earth distances are shown by the green lines with respect to all the key Earth orbit positions (summer and winter solstices, spring and autumn equinoxes) of the Earth orbit.
\begin{figure}
\includegraphics[scale=0.61]{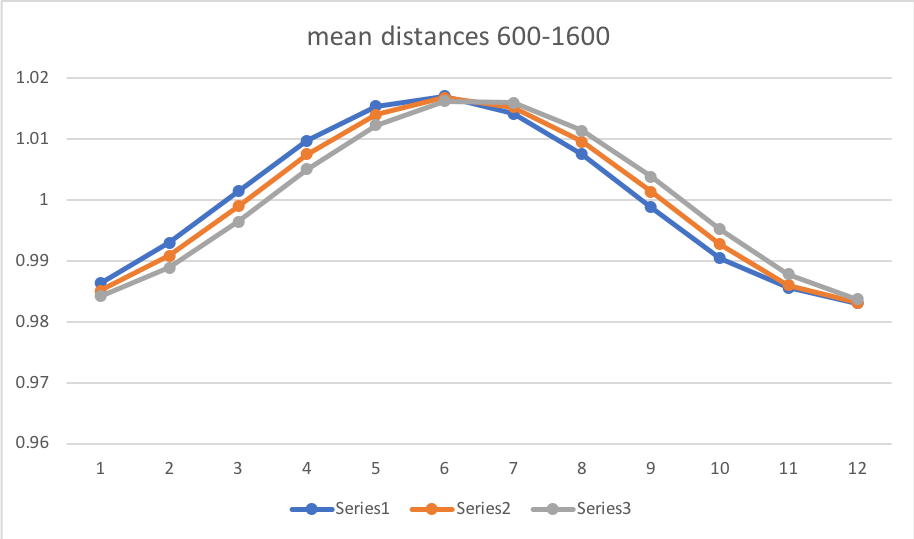}
\includegraphics[scale=0.59]{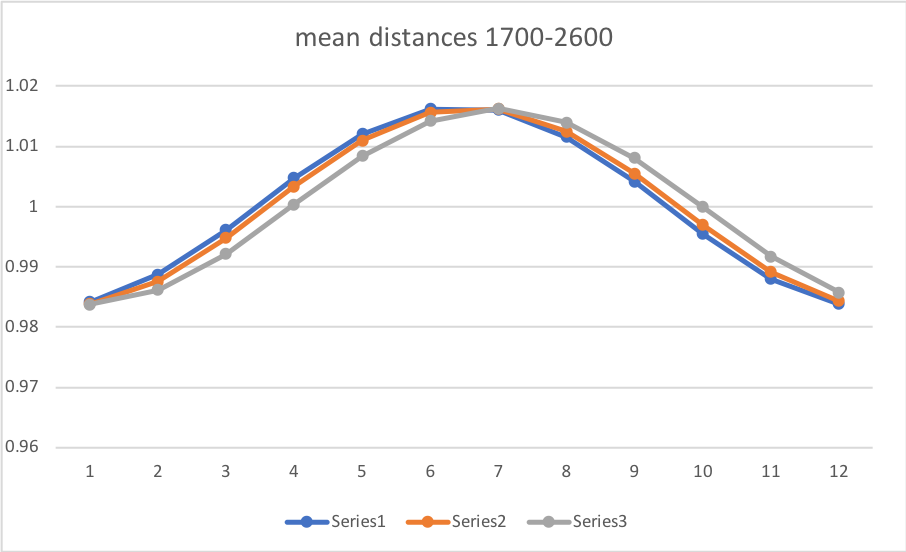}
\includegraphics[scale=0.65]{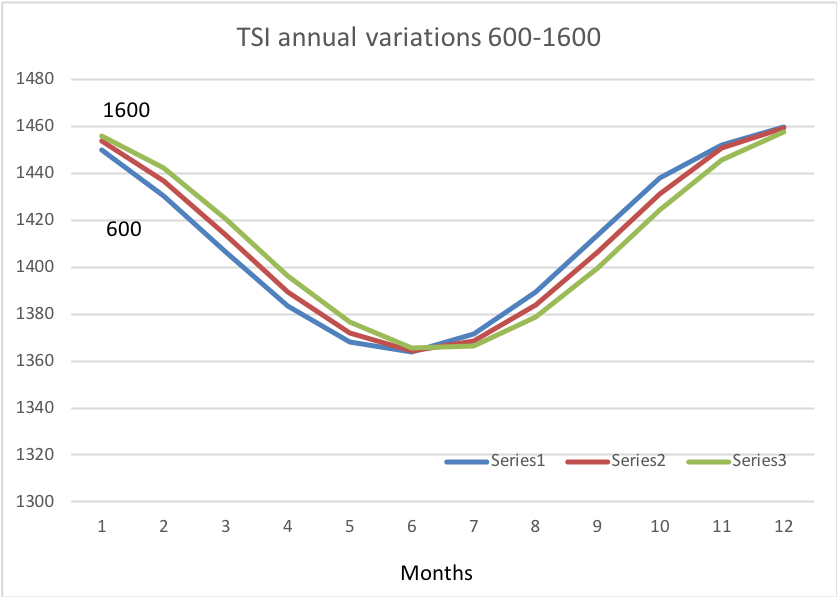}
\includegraphics[scale=0.64]{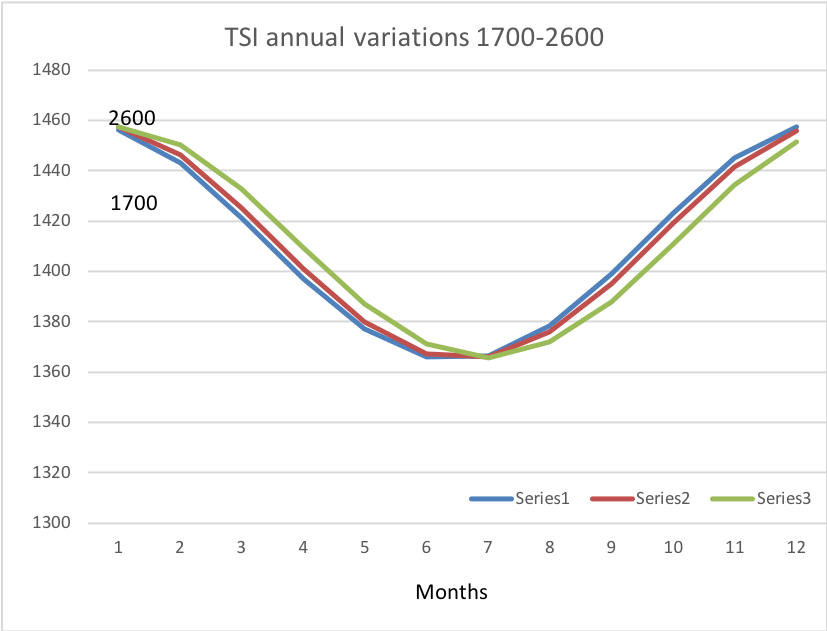}
\caption{Top row: Variations of the annual Sun-Earth distances ( axis Y, in astronomical units, au) versus months of the year (axis X) in the millennia M1 (left) and M2 (right). Left plot: blue curve - year 600, red curve -1100, and grey curve - 1600; right plot: blue curve -1700, red curve - 2020 and grey curve - 2600. Bottom row: The annual variations of TSI magnitudes (axis Y, in $W/m^2$)  in millennia M1 (600-1600) (left)  and M2 (1700-2600) (right). Axis X shows the months of a year.  All plots are courtesy of \citet{Zharkova2021}. }
\label{se_annual}
\end{figure}  
Let us present the mean monthly S-E distance variations during each sample year as plotted in Fig. \ref{se_annual}.  This, in fact, reveals that in M1 the increases/decreases of the S-E distances (left plot) are nearly symmetric over each year and centred around the summer solstice in June and winter solstice in December while in the millennium M2  the distance curve is skewed (right plot) with the maximal Sun-Earth distances being noticeably shifted in time towards mid-July for aphelion (point A in Fig.\ref{e_orbit}) and mid-January for perihelion (point P in Fig.\ref{e_orbit}). 

Namely, in M1 the local perihelion and local aphelion of the Earth elliptic orbit are shifted from the winter solstice on 21 December (point B in Fig.\ref{e_orbit}) and summer solstice on 21 June (point D in Fig.\ref{e_orbit}) forward by 5-6 days to 26-27 December  (point P in Fig.\ref{e_orbit}) and 26-28 June (point A in Fig.\ref{e_orbit}), respectively.  While in M2 the local perihelion and aphelion in 2600 are shifted even further from the elliptic orbit positions for the winter and summer solstices even further forward by 25-26 days (to 15-16 January for point P and 15-16 July for point A in Fig. \ref{e_orbit}, respectively, as seen in the right column of Sun-Earth monthly distances shown in Figs. 5  and 6 of \citet{Zharkova2021}.

 It clearly shows that the shifts in the S-E distances are reduced more in April - September and increase more in October-February of each year of millennium M2. This means that the input of solar irradiance to the Earth is not evenly distributed over the time of the Earth revolution, or over the Earth's location in the orbit. 
 
 \subsection{Variations of solar irradiance in two millennia 600-2600} \label{sec:orbit_tsi}
As established in \citet{Zharkova2021} (their Figs. 5-9), the Sun-Earth distances are changing rather differently from that defined by the simple revolution of Earth on the ellipse curve around the Sun  as Kepler's 3rd law assigns. Instead, the Sun-Earth distances are defined by two motions: the Earth and the Sun about the barycentre of the solar system. The revolution of the Sun, or the sun's wobbling, is caused by the gravitational effects of large planets of the solar system, or solar inertial motion (SIM). Therefore, the daily variations of solar irradiance over a year will be affected by the combination of the Earth's revolution on its orbit and the Sun's revolution about the barycentre. The period of the first motion  is one year, as we know, while the periods of the second motion vary from 173 to 2100 years.

In the millennium M1 (600-1600) the increase of solar irradiance during the months January-June  is nearly balanced by its decrease from July to December (Figs.11 and 12, left column  in \citet{Zharkova2021}). while in M2 the solar irradiance is noticeably higher in February-July when the atmosphere is heated by the Sun than in July-December when the atmospheric cooling occurs (see Figs. 11 and 12, right column in \citet{Zharkova2021}). This is more evident in the annual variations of the monthly averaged TSI magnitudes (Fig.\ref{se_annual}, bottom row) revealing a steady increase of the solar irradiance input in millennia M1 and, especially, in M2, during spring-summers and decrease during autumn-winters in the Northern hemisphere in each century caused by the variations of S-E distances  shown in Fig. \ref{se_annual}, top row,  discussed above. 

Hence, in the context  of two-millennial periods of the TSI variations derived in section \ref{sa_mil} from the radioisotope abundances  in terrestrial biomass, the two millennial period of oscillation of the baseline magnetic field \citep{Zharkova2019} and the total solar irradiance \citep{Zharkova2021} derived from the full disk solar magnetograms \citep{shepherd14, zhar15} have the advantage to suggest  the physical mechanism for generating  these oscillations, e.g. the  orbital effects leading to varying Sun-Earth distances caused by the solar inertial motion induced by the gravitation of large planets \citep{Zharkova2019, Zharkova2021}.  This mechanism gains more traction now with the similar suggestions proposed in the latest studies of solar activity with SSA method  \citep{lamouel2017, Courtillot2021}.

\subsection{Estimations of the TSI deposited at key points of the Earth's orbit}  \label{sec:tsi20}

 \begin{table}
\begin{center}
	\begin{tabular}{| c | c | c | c | c | c |}
	\hline
	\hline
	Date &  Distance, au &  Distance, Mkm,& TSI, $W/m^2$ &   Name \\
	\hline
5 Jan 2020 & 0.9832439 &147.091189 & 1434.583956 & perihelion 2020 \\
3 Jan 2021 &0.9832576 &147.093245 &1434.563968 & perihelion 2021 \\
4 Jan 2022 & 0.9833367 &147.105076 &1434.448571 & perihelion 2022 \\
21 Mar 2020 &0.9961576 &149.023060 &1421.448485 &spring equinox 2020 \\
21 Mar 2021 &0.9960690 &149.009803&1421.701371 & spring equinox 2021 \\
21 Mar 2022 &0.9959229 & 148.987952 &1422.118523 &spring equinox 2022 \\
21 Jun 2020 &1.0163050 &152.037062 &1365.649047 &summer solstice 2020 \\
21 Jun 2021 & 1.0162230 & 152.024802 &1365.869446 & summer solstice 2021 \\
21 Jun 2022 &1.0162025 &152.021725 &1365.924555 & summer solstice 2022 \\
4 July 2020  &1.0166939 & 152.095236 &1364.604486 &aphelion 2020 \\
6 July 2021  &1.0167292 &152.100524 &1364.509731 &aphelion 2021 \\
4 July 2022  & 1.0167151 &152.098417 &1364.547578 &aphelion 2022 \\
21 Sep 2020 & 1.0040356 & 150.201593 &1399.229673 & autumn equinox 2020 \\
21 Sep 2021& 1.0040509 &150.203877 &1404.854997 &autumn equinox 2021 \\
21 Sep 2022 &1.0041966 &150.225679 &1398.78104 & autumn equinox 2022 \\
21 Dec 2020 & 0.9837472 & 147.166485 &1433.850001 & winter solstice 2020 \\
21 Dec 2021 & 0.9837636 &147.168941 & 1433.826098 & winter solstice 2021 \\
21 Dec 2022 & 0.9838553 &147.182660 &1433.692459 & winter solstice 2022 \\
	\hline
	\end{tabular}
\end{center}
\begin{center}
	\caption{The distances and solar irradiance in $W/m^2$ at the Earth key orbital points in 2020-2022.}
	 \label{tab:tsi_mod1}
\end{center}
\end{table}

This aspect was further investigated by reference to the real  S-E data pertinent to the Earth’s orbit of the Sun and the timings associated with the Earth’s travel path between solstice and equinox at perihelion and aphelion similar to the one presented recently \citep{Fiori2022}. Here we will refer to Fig. \ref{e_orbit} showing the Earth’s schematic orbital path around the Sun.   All distances measured are on the plane of the ecliptic and at an orbital eccentricity of 0.0167.

The total solar irradiation values measured here were determined at the points of perihelion, vernal and autumnal equinoxes, and aphelion in order to obtain a percentage variation that would represent the difference in received irradiation between the northern and southern hemispheres reported from observations \citep[see, for example, ][and references therein]{Connolly2021}. Similarly to \citet{Fiori2022}, for the ephemeris of the Sun-Earth distances we compare the dates and TSI obtained by Earth when it passes through its perihelion, aphelion, vernal and autumnal equinoxes, summer and winter solstices as summarised in Table \ref{tab:tsi_mod1} for years 2020-2022.

The total solar irradiation values measured today were determined at the key orbital points of perihelion, spring, and autumn equinoxes, and aphelion. Note that the current TSI calculations were normalized on the TSI of 1366 $W/m^2$ restored for the aphelion of the orbit in 1700 \citep{Lean1995}. It is evident that the S-E distances, and thus TSI, change significantly depending onthe position of the Earth on its elliptic orbit, leading to the change of seasons on Earth in different hemispheres.

The TSI received during the time from the solstice to an equivalent time past the perihelion is directed to the southern hemisphere that is experiencing summer. This time is winter in the northern hemisphere. The reverse occurs at aphelion. The difference in the timing of the summer of the southern hemisphere to the timing of the summer of the northern hemisphere is caused by the Earth’s orbital velocity at the respective times. It takes longer for the earth to travel the same distance on the circumference of the ellipse during aphelion than it does during perihelion. The Earth lingers longer in the reduced irradiation from the sun at aphelion and receives about 4.7$\%$ more irradiation than the earth at perihelion, principally directed to the northern hemisphere \citep[in Table 5.2 in ][]{Fiori2022}.

However,  these S-E distance and TSI variations are not the same for different years (2020 to 2022 as examples). it can be seen from Table \ref{tab:tsi_mod1} that the S-E distances at perihelion and autumn equinox are steadily increasing from 2020 to 2022 while at aphelion and spring equinox the S-E distances are decreasing with every year because the Sun is moving towards the spring equinox owing to SIM, or the large planet gravitation \citep{perminov2018}. This change of S-E distances at the key points during 2020-2022, in turn, leads to the increase of TSI at Earth at the northern spring equinox and aphelion and decrease at the northern autumn equinox and perihelion. These differences between TSI keep increasing from 2020 to 2022 and during the further few centuries until at least 2500 as shown above. 

Then we estimate the number of days spent by Earth between the key points of its orbit in 2020 compared to 2021 as described by \citet{Fiori2022} during the period of time when the Earth travels from the solstice to the perihelion or the aphelion and away from the perihelion or aphelion. In 2020 the time spent by the Earth between the autumn and spring equinox was 182 days and between the spring and autumn equinox was 184 days. While in 2021 these numbers become 181 and 184, respectively  \citep{Fiori2022}.  

 The amount of solar irradiance deposited on Earth in 2020 between the autumn and spring equinox when there is a summer in the Southern hemisphere and winter in the Northern one is 256935 $W/m^3$, or 49.897$\%$ of the TSI while the amount of solar irradiance deposited between the spring and autumn equinox when summer is in Northern and winter is in Southern hemispheres reaches 257995 $W/m^3$, that constitutes 50.103$\%$ of overall TSI. This distribution is rather close to the estimation of 49.85$\%$ and 50.15$\%$ provided by \citet{Fiori2022} considering simplifications in their approximate method. Although, based on the findings in section \ref{sec:orbit_tsi}, this difference between the TSI in the opposite hemispheres is expected to increase during the next few centuries until 2500 because of a further change in the official S-E distance ephemeris supposedly caused by SIM.

A further consideration is a} speed, at which the angle of obliquity changes as the Earth moves away from the summer and winter solstices as described by \citet{Fiori2022} (see Figure 5.7).
The rate of change of return of the angle of obliquity from obstructing the solar irradiance at the winter solstice to the autumn equinox is smaller than the rate of change of return of the angle of obliquity from the summer solstice to the spring equinox. Thus, the angle of obliquity changes at a rate 3.8$\%$ slower at aphelion \citep{Fiori2022}. Hence the area of the Northern hemisphere is receiving solar radiation longer at aphelion than at perihelion.

Taking into consideration the characteristics of the properties and rates of the S-E distances and TSI changes in 2020 or 2021 as described here and by \citet{Fiori2022} (see their Figures 5.5, 5.6, and 5.7), the one can deduce that summers in the Northern hemisphere last longer and receives a greater TSI than the Southern hemisphere’s summer whilst there is a longer winter in Southern hemisphere, which receives a smaller amount of solar radiation than during the Northern hemisphere’s winter. This can explain why the polar region of the Southern hemisphere should receive less solar radiation than the polar region of the Northern hemisphere. 

\citet{Fiori2022} suggested that this difference can lead to a greater probability of the Southern polar region entering into ice age conditions in the future faster than the Northern hemisphere. However, these changes could be seen on much larger timescales than the two-millennia Hallstatt's cycle discussed in the current paper.

\subsection{Differences in the TSI depositions to Earth in the two millennia} \label{sec:tsi_mill}
From the daily magnitudes of TSI shown in Figs. 11 and 12 of  \citet{Zharkova2021}, it is possible to count the real distances between the Sun and Earth obtained from the official ephemeris and the total annual amount of TSI emitted by the Sun towards the Earth in each year of both millennia. If this amount does not change from year to year, then TSI is, indeed, the same for each year for both centuries, as currently assumed. 

However, because in the millennia M1 (600-1600) and M2 (1600-2600) there is the shift of the maximum points of the Sun-Earth distances (Fig.\ref{se_annual}, top plots), or the minimum point of the TSI annual variations (Fig.\ref{se_annual}, bottom plots) from 21 June (M1) to 15-16 July (end of M2), a detectable difference occur between the annual TSI inputs in M1 and M2.  

The annual magnitudes of total solar radiation (TSI)  deposited to the Earth during the sampled three years (600, 1100, 1600, and 1700, 2020, 2600) in the two millennia were shown in Fig.14 \citet{Zharkova2021}. These were calculated for the cases: a)  by adding together the averaged monthly TSI magnitudes (left plot) calculated for the S-E distances shown in Fig. \ref{se_annual}, top plots, when only 12 magnitudes per year (for 12 months) are  considered;  b) by adding together the daily TSI  magnitudes (right plot) taken from Figs. 11  and 12  of  \citet{Zharkova2021}, associated with the daily magnitudes of TSI  (for 366 days for the leap years). 

These two plots clearly demonstrated \citep{Zharkova2021} that the monthly TSI variations (case a) show the increase of TSI by about 1-1.3 $W/m^2$ in 2020 compared to 1700. This TSI increase found from the S-E distance ephemeris is close to the magnitude of 1-1.5 $W/m^2$ reported from the current TSI observations \citep{krivova11}. However,  it was shown that the annual TSI magnitudes, calculated from the daily S-E distances (case b)  reveal much larger annual increases of the TSI by about 20-25 $W/m^2$ ($>$1.8$\%$) in M2 than in millennium M1.

This indicates that the averaged TSI will increase in M2 by about  2.5-2.8 $W/m^2$ per century, or  (0.18-0.20)$\%$, compared to the TSI in 1700. This calculation uncovers a hidden TSI input present in millennium M2 (1600-2600), which is likely caused by the gravitational effects of large planets, or solar inertial motion. This effect is significantly underestimated if only the averaged monthly TSI magnitudes are used  (compare here Fig. \ref{se_annual}, bottom left and right plots)  as well as Fig. 14, top and bottom plots,  in \citet{Zharkova2021}). The important issue is how this extra solar radiation is distributed between the hemispheres during the Earth revolution about the barycentre or the level of exposure of the Earth to solar radiation during different seasons and different positions of the Earth orbit parameters \citep{milankovich98, steel2013}. 
 
Furthermore, It was shown that the distribution of the TSI to the opposite hemispheres varies during a year. At the start of any year when the Earth is turned to the Sun with its Southern hemisphere, any decrease and increase of solar radiation occurring during this time are mostly absorbed by the Southern hemisphere. When the Earth's orbit approaches the spring equinox in March, the distribution of TSII between the hemispheres becomes nearly even, while in April-July the main part of the solar radiation input is shifted toward the Northern hemisphere. Theoretically (by Kepler’s law), the Earth aphelion should occur on the summer solstice, 21 June, while, in reality, it is shifted to 5 July in 2020 and to 16 July in 2500. 

Hence, because of the shift of the Sun towards the spring equinox position of the Earth orbit owing to the gravitation of large planets, in the millennium M2 the Northern hemisphere should get the extra solar radiation not only in the first six months of a year but especially, in the 25 days between 21 June and 16 July when the aphelion is approached in 2500. This extra TSI deposited to the Earth for the Northern hemispheres in the spring and summer is not compensated later in Autumn and winter. This is possibly affecting to some extent the Southern hemisphere as well because of extra cooling occurring because of a shift of the local perihelion to 16 January in 2500. 

Therefore, the deposition of TSI in the millennium M2 can be essentially higher than in millennium M1. This can lead to different seasonal changes in the opposite hemispheres because of the uneven shifts of Sun-Earth distances on the orbit owing to SIM. This extra input of the TSI amount caused by SIM (from the variations of a distance $d$ in inverse squares formula for solar radiation) will undoubtedly add to the magnitude of solar heating coming from the solar activity itself as shown in Fig. 4 (bottom plot, blue lines) in \citet{Zharkova2021}.  This, in turn, can  account for a large solar input to  the terrestrial temperature increase shown by the red curves  in Fig. 4, bottom plot, red lines) in \citet{Zharkova2021}, in addition to any other natural and anthropogenic causes considered so far. 

 \section{Discussion and conclusions} \label{conc}
 \subsection{Derived periods and their links to solar activity and orbital motion}
 In this paper we carried out spectral analysis with the wavelet and Fourier transforms for the averaged sunspot numbers representing the toroidal magnetic field of the solar dynamo and the summary curve derived from the two largest eigenvectors derived from solar background magnetic field representing a poloidal magnetic field of the solar dynamo. Despite differences between the proxies of solar activity used, we established that the key period of solar activity is 11-years (or 10.7-years to be precise)  and it is the same for both proxies. There is also a period of about 22 years, or 21.4 years to be precise, detected from the summary curve of eigenvectors of SBMF where the polarity of the magnetic field is taken into account.  These periods are derived from the sunspot numbers, summary, and modulus summary curves of eigenvectors of magnetic waves generated by dipole magnetic sources.
 
 There was another period detected in the sunspot data of about 101 years, or the Gleissberg's cycle, which does not appear in the modulus summary curve of the two largest eigenvectors generated by dipole magnetic sources.  This is understood during the interpretation of the observed activity when Gleissberg's cycle can be accounted for from consideration of the next pair of eigenvectors generated by quadruple magnetic sources. This finding allows us to confirm the conclusion of many other authors that the key period of solar activity is a 22-year cycle \citep{Cameron2017, Cameron2019}.
 
 Another important period detected from the summary and modulus curves of eigenvectors of dipole magnetic waves in the 2000 years (1200-2300) was the period of 344 years, which averages the five periods ranging from 330 to 380 years, constituting give grand solar cycles separated by grand solar minima (GSMs). This grand period of 330-350 years for a grand solar period is likely caused by the wave interference (beating effect) of the two largest eigenvectors. The same grand period of 350 years is also recorded in the wavelet and  Fourier spectra of solar irradiance in the past 12 000 years. This grand period is well recorded in the past observations of solar activity via numerous GSMs: Maunder, Wolf, Oort, Homeric, and other grand solar minima.
 
The fact that eigenvectors come in pairs assumes that these waves are formed in different layers of the solar interior, which have different properties and thus produce the dynamo waves with the close but not equal frequencies (or periods) and the phase difference between these two waves varying in time as a cosine function. We explored the opportunity of these dynamo waves to be formed by the dynamo action in two layers with slightly different velocities of meridional circulation that induces the beating effect during the wave interference and forms these grand solar cycles like those beats appearing when tuning a piano with a fork. 

Furthermore, because these two dynamo waves are formed in different layers (one in the solar interior near the bottom of the solar convective zone and another close to the shallow layer beneath the surface, this allows us to reconcile the existing dynamo models considering some stochasticity of the waves in the shallow layer caused by its properties and another stochasticity appearing naturally from the phase difference between the waves that can explain essential features of the solar dynamo waves.

The other, larger, period, which was reported from the observation of solar radiations recorded in the abundances of the isotope of carbon 14 in the terrestrial biomass, was the two millennial period of 2200-2300 years which also coincided with the period of oscillations of the magnetic field baseline derived directly from the summary curve.  Since none of the dynamo models can account for any oscillations of the zero line of a magnetic field, we looked at the cause of this oscillation in the orbits of the solar planets and the sun itself, shifting from the ellipse focus towards the Earth's orbit points (currently, towards the spring equinox point), so-called solar inertial motion. This SIM causes standard deposition of solar radiation in one millennium (600-1600)  and much stronger deposition of solar radiations in the other millennium (1600-2600) independently on solar activity processes in the Sun itself.

By comparing all these periods together one can understand better the effects of not only solar activity but also the interaction of the Sun with the planets of the Solar System and their effects on the terrestrial temperature.

\begin{figure*}
\includegraphics[scale=0.73]{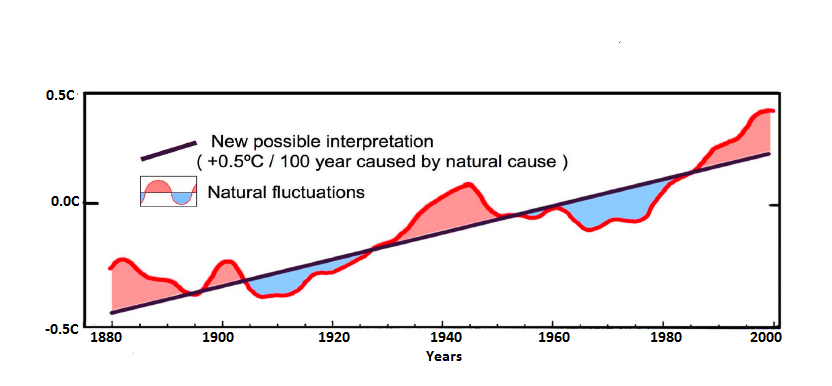}
\caption{ Terrestrial temperature variations from 1880 until 2000 showing the increase of the baseline temperature (solid black line over imposed by the decadal oscillations shown by the blue (decreasing) and red (increasing) shapes.  A courtesy of \citet{akasofu2010}. } 
\label{temp}
\end{figure*}

\subsection{TSI variations and terrestrial temperature } \label{sec:tsi_temp}
Let us try to evaluate how these variations of solar irradiance can affect terrestrial temperature from the general similarity approach using the plot presented in Fig. \ref{temp}, courtesy of \citet{akasofu2010}. The variations of the temperature have a decadal trend linked to 11-year cycles and the baseline variations linked to other mechanisms of solar irradiance variations (grand solar minima \citep{Eddy1976, zhar15}, solar inertial motion \citep{charvatova1988, Scafetta2016, Zharkova2021}, other orbital variations caused by Milankovich cycles \citep{milankovich98}.

The TSI variations caused by the solar activity in normal cycles of the 11 years and during grand solar minima (similar to the Maunder Minimum, MM) can be described as follows.
\begin{enumerate}
\item Solar irradiance $S$ variations at Earth owing to 11 year cycle is about  0.1$\%$ of the average magnitude  of TSI $S$ (1366 W$m^{-2}$ accepted in this study) increasing by 1.4 W$m^{-2}$ during maxima and decreasing during minima \citep{Lee1995, willson1991}. These estimations are also supported by the other researchers  \citep{Lockwood1999, Fligge2000} showing sometimes up to 0.4$\%$ contributions of active regions into the solar radiance intensity $I_{\bigodot}$. 

The terrestrial temperature variations during 11-year cycle are small but noticeable as shown by decadal variations in Fig.\ref{temp} being proportional to the averaged sunspot numbers \citep{Eddy1976, Foukal2006, Stauning2011} and many terrestrial factors like variations of ocean temperature, directions off jets and ENSO variations with El Nino and La Nina events \citep{Connolly2021}. This leads to recognizable increases of the terrestrial temperature during or close to the maxima of solar activity and decreases during or close to the minima.
\item Solar irradiance $S$ variations (decreases) at Earth owing to GSM is about 2.5-3 $W/m^2$, or 0.22 $\%$ of $S$ \citep{Lean1995, Lean2000, Fligge2000} as shown in Fig.2  (top plot) in \citet{Zharkova2021}.  This would lead to recognisable decreases of the baseline temperatures during GSMs, similar to that reported for the Maunder Minimum \citep{Shindell2001, Miller2012, Easterbrook16}.
 \end{enumerate}

 The terrestrial temperature curve averaged from the fluctuations in 11-year cycles presented in Fig.2  (bottom plot) in \citet{Zharkova2021} shows a reduction during the MM of the average terrestrial temperature by about 1.0$^\circ$C \citep{Shindell2001, Miller2012, Easterbrook16}, e.g. the decrease of TSI by 0.11$\%$ secures a decrease of the terrestrial temperature by approximately 0.5$^\circ$C.  

\subsection{Effects of the upcoming Grand Solar Minimum (2020-2053)} \label{mod_gsm}

In the next 30 years, the Sun is entering a period of reduced solar activity, the modern grand solar minimum \citep{zhar15}, which can be called a ‘mini ice age’, similar to the Maunder Minimum.  This is already confirmed by the sharpest increase of spotless days in cycle 25 compared to all other cycles \citep{SILSO, Zharkova2022},  or by a sharp increase of the methane index in the terrestrial atmosphere supposedly caused by a decrease of the UV radiation of the Sun igniting a decrease of the terrestrial ozone abundances caused by it \citep{Archibald2022}. There were also signs of the reducing terrestrial temperature recorded in the past decades until the minimum between cycles 24 and 25 \citep{Held2013, Kosaka2013, Connolly2021, Alimonti2022}, which now are overridden by an increase of this temperature that is likely \citep{Stauning2011} caused by the approaching maximum of cycle 25 combined with the effect of the solar inertial motion on the baseline temperature \citep{akasofu2010, Zharkova2021}. 

Given that cycle 25 is still at its maximum, further confirmation ofthe reduction of solar activity and its effect on the terrestrial atmosphere will come during the next decade, in the descending phase of cycle 25  and the solar minimum between cycles 25 and 26. This is when the full grand solar minimum will occur in the Sun leading to a significantly reduced solar magnetic field on the solar surface caused by the disruptive interference of two magnetic waves generated by the double dynamo in the solar interior  \citep{zhar15}. Since the SIM effects have a  long period of 2200-2300 years  \citep{Scafetta2016, Zharkova2019, Zharkova2021}, then, it will not change much a deposition of solar radiation to the Earth within 30 years of GSM, it will be mainly defined by the GSM. The first modern GSM1 occurs in 2020–2053 \citep{zhar15, zhar2020} and  the second modern GSM2 will happen in 2370–2415 \citep{zhar15, zhar2020}.  

  Because solar irradiance andthe terrestrial temperature already increased since the MM as is clearly recorded from the terrestrial temperature variations \citep{akasofu2010, Stauning2011},  the terrestrial temperature during the first modern GSM1 is expected to drop by about 1.0$^\circ$C  to become just  0.5$^\circ$ C higher than it was in 1700. The second GSM2  will arrive in 2375-2415 \citep{zhar15} and because at this time the Sun will move closer towards the Earth owing to SIM \citep{Zharkova2019, Zharkova2021},  the starting terrestrial temperature before GSM2 will be higher by 1.5$^\circ$C  than at the start of the current GSM1. The reduction of TSI caused by solar activity in GSM2 will lead to a reduction of the terrestrial temperature by 1.0$^\circ$C, leaving the Earth warner by about 0.5$^\circ$C during the GSM2 compared to the temperature at the end of  the current GSM1, or by 1.0$^\circ$C higher than in 1700. 

 Therefore, the periods of solar activity of 10.7, 21.5, 350, and 2100-2300 years derived from the variations of averaged sunspot numbers, eigenvectors of  SBMF and solar irradiance linked to isotope abundances can jointly help us to understand better the solar activity on short and long timescales, provide some plausible logical interpretation and anticipate their expected effects on the terrestrial environment.




\section*{Acknowledgments}
The authors would like to thank the Solar Influences Data Analysis Center (SIDC) at the Royal Observatory of Belgium for providing the corrected averaged sunspot numbers. The authors also express their deepest gratitude to the staff and directorate of Wilcox Solar Observatory for providing the coherent long-term observations of full disk synoptic maps of the solar background magnetic field. All the authors wish also to acknowledge the support  of the ZVS Research Enterprise Ltd., UK.


\section*{Data Availability Statement}
The datasets [ANALYZED] for this study can be found in the solargsm repository] [https://solargsm/depository].
\bibliographystyle{aasjournal}
\bibliography{zhark_cycles}
\end{document}